\documentclass[14pt]{article}
\usepackage[a4paper]{geometry}
\usepackage{slashed,verbatim}
\usepackage[numbers,sort&compress]{natbib}
\usepackage{amsmath}

\newcommand{\Rmnum}[1]{\expandafter\@slowromancap\romannumeral #1@}
\makeatother
\usepackage{amssymb}
\usepackage{float}
\usepackage{comment}
\usepackage{appendix}
\usepackage{graphicx}
\usepackage{subcaption}
\usepackage{dcolumn}
\usepackage{bm}

\usepackage[colorlinks=true,linktocpage=true,
linkcolor=blue,citecolor=blue]{hyperref}
\newcommand{\ba}{\begin{eqnarray}}
\newcommand{\ea}{\end{eqnarray}}

\usepackage{bm}
\usepackage{authblk}

\begin{document}
		\large
		\title{\bf{Thermoelectric response of a hot and weakly magnetized anisotropic QCD medium}}

 \author[1, 2]{Salman Ahamad Khan\footnote{skhan@ph.iitr.ac.in}}
 \author[1,3]{Debarshi Dey\footnote{ddey@ph.iitr.ac.in}}

\author[1]{Binoy Krishna
			Patra\footnote{binoy@ph.iitr.ac.in}}
			
			\affil[1]{Department of Physics,
			Indian Institute of Technology Roorkee, Roorkee 247667, India}
			
			\affil[2]{Department of Physics, 
Integral University, Lucknow 226026, India}

			\affil[3]{Department of Physics, Indian Institute of Technology Bombay, Mumbai 400076, India}

		\maketitle
	\begin{abstract} 
		We have studied the Seebeck and Nernst coefficients of 
		a weakly magnetized hot QCD medium having a weak momentum anisotropy within the kinetic theory approach.
		The thermal medium effects have been incorporated in the framework
		of a quasi-particle model where the medium dependent
		mass of the quark has been calculated using perturbative
		thermal QCD in the presence of a weak magnetic field which leads to different masses for the left ($L$) and right ($R$) handed chiral quark modes. We have found that the Seebeck and Nernst coefficient magnitudes
		for the individual quark flavors as well as for the composite medium are decreasing functions of temperature and decreasing functions of anisotropy strength. The Nernst coefficient magnitudes are about an order of magnitude smaller than their Seebeck counterparts, indicating the Seebeck effect constitutes a stronger response than the Nernst effect. 
The average percentage change corresponding to switching between quasiparticle modes ($L\to R$ or $R\to L$) is an order of magnitude smaller for Nernst coefficients, compared to the Seebeck coefficients.

	\end{abstract}
	
	\section{Introduction}
	Heavy ion collisions at ultra-relativistic energies give rise to a state of matter comprising of asymptotically free quarks and gluons-the Quark Gluon Plasma (QGP). Several observables are considered as signatures of creation of such a medium; this includes photon and dilepton spectra\cite{Shuryak:PLA757'1978,Kapusta:PRD44'1991}, quarkonium suppression\cite{Blaizot:PRL77'1996,Satz:NPA783'2007,Rapp:PPNP65'2010}, elliptic flow\cite{Bhalerao:PLB641'2006,Voloshin:PLB659'2008}, jet quenching\cite{Wang:PRL68'1992,Adcox:PRL88'2001,Chatrchyan:PRC84'2011}, etc. 
	Experimentally, significant evidence now exists of the observation of these signals, and thus, of the creation of QGP matter in
	Ultrarelativistic Heavy Ion collisions (URHICs) at experimental facilities such as the Brookhaven National Laboratory
	Relativistic Heavy Ion Collider (RHIC) \cite{Arsene:NuclPhysA757'2005,Adams:NuclPhysA757'2005,Adcox:NuclPhysA757'2005} and Large Hadron Collider (LHC)\cite{Carminati:JPhysG30'2004,Alessandro:JPhysG30'2006}.  
	Right after the formation of QGP, it expands and cools and transitions into a mildly interacting collection of hadrons. 
	At very small baryon chemical potentials ($\mu_B \approx 0$), with massive quarks, the results of lattice QCD indicate that the transition is actually an
	analytic crossover rather than a true phase transition. \cite{Aoki:PLB643'2006,Borsanyi:JHEP1009'2010,Borsanyi:PRD92'2015,Ding:IntJModPhysE24'2015}. The sign problem of lattice QCD at finite $\mu_B$ makes it an unreliable tool to explore large parts of the QCD phase diagram\cite{Hands:NuclPhysB106-107'2002,Alford:NuclPhysProcSuppl117'2003}.
	
	The QGP lives for a very short period of time and does not expand at the same rate in all the directions. The colliding nuclei are Lorentz contracted due to their relativistic speeds. The overlap volume of such nuclei in a non-central collision is anisotropic in the plane perpendicular to the beam (transverse plane). This spatial asymmetry in coordinate space gets converted into an opposite asymmetry in the momentum space which is ultimately reflected in the hadron $p_T$ spectra. A convenient way of taking into account the anisotropy of the medium was introduced by Romatschke and Strickland wherein the anisotropic distribution function is obtained from an arbitrary isotropic one by the rescaling of only one direction in
	momentum space, \emph{i.e.} by stretching or squeezing the isotropic distribution function of the medium constituents (partons)\cite{Romatschke:PRD68'2003}. This parametrisation of the anisotropy involving a single direction and a single anisotropy parameter, $\xi$ (spheroidal momentum anisotropy), was used to calculate the collective modes of finite temperature QCD and study their impact in thermalization of the QGP medium\cite{Romatschke:PRD68'2003,Romatschke:PRD70'2004,Carrington:PRC90'2014,Schenke:PRD74'2006}. Hard loop effective theories have also been used to study anisotropy; its equivalence with the kinetic theory approach was shown by Mr\'{o}wczy\'{n}ski and Thoma, wherein they calculated the self-energies and dispersion relations for
	QGP partons\cite{Thoma:PRD62'2000}. Existence of instabilities associated with gluon collective modes in an anisotropic QGP has been observed and their growth rates have also been calculated\cite{Randrup:PRC68'2003,Strickland:PREP682'2017}. This description has also been used to study photon and dilepton production from the
	QGP \cite{Schenke:PRD76'2007,Bhattacharya:PRD93'2016}, the QGP heavy quark potential \cite{Nopoush:JHEP09'2017}, and
	bottomonia suppression\cite{Krouppa:UNI2'2016}.  In addition to the case of spheroidal anisotropy, ellipsoidal momentum anisotropy has also been considered, which is characterised by two or more independent anisotropic parameters. In particular, parton self energies have been calculated in an ellipsoidally anisotropic QGP\cite{Kasmaei:PRD97'2018}. Anisotropic momentum distributions have also been used in relativistic hydrodynamic models to study the evolution of QGP to hadrons\cite{Florkowski:PRC83'2011,Martinez:NPA848'2010,Alqahtani:PPNP101'2018,Alqahtani:PRL119'2017,Alqahtani:PRC96'2017}. In fact, anisotropic hydrodynamics (aHydro) has been more accurate than its isotropic counterpart in the description of non-equilibrium dynamics\cite{Florkowski:PRC89'2014,Nopoush:PRC90'2014,Tinti:NPA946'2016,Strickland:NPA956'2016,Strickland:PRD97'2018,Strickland:JHEP12'2018}. As such, transport coefficients like electrical conductivity\cite{Rath:PRD100'2019}, heavy quark drag and diffusion coefficients\cite{Kumar:PRC105'2022} have also been evaluated in an anisotropic plasma.  
	
	Apart from causing an anisotropic expansion of the created matter, non-central heavy ion collisions also lead to creation of large magnetic fields\cite{Tuchin:AdvHEP'2013}. These magnetic fields, produced mainly by the spectator protons moving away from each other at relativistic speeds, reach magnitudes upto $eB\approx10^{-1}
	m_{\pi}^2\,(\approx 10^{17}$ Gauss) for SPS 
	energies,  $eB\approx m_{\pi}^2$ for RHIC energies 
	and $eB\approx 15m_{\pi}^2$ for LHC energies\cite{Sokov:IntJModPhysA24'2009}. The decay rate of the magnetic field depends strongly on the electrical conductivity of the medium which is exposed to the field\cite{Tuchin:PRC82'2010,Tuchin:PRC83'2011,Marty:PRC88'2013,Ding:PRD83'2011,Gupta:PLB597'2004,Amato:PRL111'2013,Aarts:PRL99'2007,Puglisi:PRD90'2014,Greif:PRD90'2014,Hattori:PRD96'2017}. Depending on the strength of the background magnetic field, several interesting phenomena of the created matter can be probed. A strong background magnetic field causes separation of charges in a chiral QGP medium leading to a magnetic field dependent current, which is non-Maxwellian and has no analog in classical physics. This is the Chiral Magnetic effect\cite{Kharzeev:Nucl.Phys.A803'2008,Fukushima:PRD78'2008,Kharzeev:Ann.Phys.325'2010}. Other phenomena induced by strong magnetic fields include magnetic catalysis\cite{Shovkovy:LecNotesPhys871'2013}, chiral magnetic wave\cite{Kharzeev:PRD83'2011}, axial magnetic effect\cite{Braguta:PRD89'2014,Chernodub:PRB89'2014}, etc. A small electrical conductivity, however, would lead to only a small fraction of the initial magnetic field surviving when the created matter thermalizes. This has motivated several studies where the background magnetic field is considered to be weak. Further, such a weak field can give rise to novel phenomenological results like the lifting of mass degeneracy between the left and right handed quark effective masses\cite{Das:PRD97'2018}. Transport phenomena in the presence of weak magnetic field has been under investigation in the recent past\cite{Panday:PRD105'2022,Das:PRD101'2020,Thakur:PRD100'2019,Feng:PRD96'2017,Rath:EPJC82'2022,Shaikh}

In this article, we study for the first time, the thermoelectric response of an anisotropic QGP medium in the presence of a weak magnetic field, by incorporating the non-degenerate left and right chiral quasiparticle masses of quarks. Fluctuations in the initial energy density of heavy-ion collisions can create large temperature differences between the central and peripheral regions of the fireball\cite{Schenke:PRL108'2012}. This, coupled with a finite chemical potential can potentially give rise to thermoelectric phenomena: Seebeck and Nernst effects. The ability to convert a temperature gradient into an electric field is quantified by the Seebeck and Nernst coefficients. Thermoelectric phenomena have been previously investigated both in the presence and absence of magnetic field in a thermal QCD medium\cite{Bhatt:PRD99'2018,Das:PRD102'2020,Dey:PRD102'2020,Dey:PRD104'2021,Kurian:PRD103'2021,Zhang:EPJC81'2021,Khan:arxiv}.

In what follows, we present the calculation of the thermoelectric coefficients of a QGP medium, taking into account the expansion induced anisotropy of the momentum distribution of the partons, in the presence of a weak background magnetic field. The paper is organized as follows: In section II the quasiparticle model used in this work is described. In section III, the calculation of Seebeck and Nernst coefficients are outlined both for single component and multi component mediums. In section IV, the results are plotted and their interpretations discussed. Finally, we conclude in section V.

\section{Quasiparticle model}\label{two}
The central 
feature of quasi particle models is that a strongly
interacting system of massless quarks and gluons 
can be described in terms of 
 massive, weakly interacting quasi-particles 
originating due to the collective excitations. There are many 
quasi-particle models 
such as Nambu-Jona-Lasinio (NJL) 
model and PNJL model 
\cite{Fukushima:PLB591'2004,
	Ghosh:PRD73'2006,Abuki:PLB676'2009,Tsai:JPG36'2009}
which are based on the 
respective effective QCD models, effective fugacity model
\cite{Chandra:PRC76'2007} and the 
recently proposed quasi-particle model based on the 
Gribov–Zwanziger quantization
~\cite{Su:PRL114'2015,Florkowski:PRC94'2016,
	Jaiswal:PLB11'2020}. 
In this study, we have used quasi-particle 
model~\cite{Bannur:JHEP09'2007} which has only a single adjustable
parameter and the medium effects enter through 
the dispersion relations of the 
quark and gluon quasi-particles. The temperature
and magnetic field-dependent masses of the 
quarks and gluons 
have been computed from the poles of their
resummed propagators obtained from Dyson-Schwinger equations.
The respective self-energies have been calculated using perturbative thermal 
QCD in a strong magnetic field background.
The quasi particle mass of the $i^{th}$ flavor
is written phenomenologically
as~\cite{Bannur:JHEP09'2007} 
\begin{eqnarray}
m_i^2=m_{i,0}^2+\sqrt{2}m_{i,0}m_{i,T}+m_{i,T}^2,
\label{para_massT}
\end{eqnarray}
where $m_{i,0}$ and $m_{i,T}$ are the 
current quark mass and medium generated quark mass. $m_{i,T}$  
has been calculated using the HTL perturbation theory as \cite{Braaten:PRD45'1992,Peshier:PRD66'2002}
\begin{eqnarray}
m_{iT}^2=\frac{g'^2T^2}{6}\left(1+\frac{\mu^2}{\pi^2 T^2}\right),
\label{quarkmassT}
\end{eqnarray}
where $g'=\sqrt{4\pi\alpha_s}$ refers to the coupling
constant which depends on the 
temperature as
\begin{eqnarray}
\alpha_s (T)=\frac{g'^{2}}{4 \pi}=
\frac{6\pi}{(33-2N_f) \ln \left(\frac{Q}
	{\Lambda_{QCD}}\right)},
\label{coupling_T}
\end{eqnarray}
where, $Q$ is set at $2 \pi\sqrt{T^2+\frac{\mu^2}{\pi^2}}$.\par

In the presence of a strong magnetic field, the coupling constant $g=\sqrt{4\pi\alpha_s}$ depends on the 
temperature, chemical potential and magnetic field as~
\cite{Ayala:PRD98'2018} 
\begin{eqnarray}
\alpha_s(\Lambda^2,eB) =\frac{g^2}{4\pi}
=\frac{\alpha_s(\Lambda^2)}{1+
	b_1\alpha_s(\Lambda^2)\ln\left(\frac{\Lambda^2}
	{\Lambda^2+eB}\right)},
\label{alpha_B}
\end{eqnarray}
with 
\begin{eqnarray}
\alpha_s(\Lambda^2)=\frac{1}{
	b_1\ln\left(\frac{\Lambda^2}
	{\Lambda_{\overline{MS}}^2}\right)},
\end{eqnarray}
where $\Lambda$ is set at $2\pi \sqrt{T^2+\frac{\mu^2}{\pi^2}}$ for quarks,
$b_1=\frac{11N_c-2N_f}{12\pi}$ and
$\Lambda_{\overline{MS}}=0.176$ GeV.\\

{Similarly the effective  quark mass for $i^{th}$ 
	flavor in the case of a weak magnetic field can 
	be parameterized like in
	the earlier cases as
	\begin{eqnarray}
	m^2_{i,w} = m_{i0}^2 + \sqrt{2}m_{i0}m_{i,L/R}+m_{i,L/R}^2,
	\end{eqnarray}
	where  $m_{i,L/R}$ refers to the 
	thermal mass for the left- or right-handed 
	chiral mode of  $i^{th}$ flavor  
	which can be 
	evaluated from the  Dyson-Schwinger equation
	in weak magnetic field 
	\begin{eqnarray}\label{inverse_prop}
	S^{*-1}(P) &=& 
	\slashed{P} - \Sigma(P).
	\end{eqnarray}
	$\Sigma(P)$ is the quark self energy in
	the weakly magnetized thermal medium.
	The general  structure of 
	quark self energy in the covariant form 
	at finite temperature and magnetic field can 
	be written as \cite{Das:PRD97'2018}
	\begin{eqnarray}\label{52}
	\Sigma(P) = -A'\slashed{P}-
	B'\slashed{u}-C'\gamma_5\slashed{u}-
	D'\gamma_5\slashed{b},
	\label{self_weak}
	\end{eqnarray} 
	where  $u^{\mu}=(1,0,0,0)$ and $b^{\mu}=(0,0,0,1)$ 
	refer to the velocity of the heat bath and the direction of the magnetic field,
	respectively. $A', B', C', D'$ 
	are the structure functions which can be 
	evaluated by taking the appropriate contractions of 
	Eqn.~\eqref{52} as~\cite{Das:PRD97'2018} 
	\begin{eqnarray}\label{A}
	A'\left(p_0, p_{\perp}, p_z\right)&=&
	\frac{1}{4}\frac{\text{Tr}(\Sigma(P)\slashed{P})
		-(P.u)\text{Tr}(\Sigma(P)\slashed{u})}{(P.u)^2-P^2},\\ \label{B}
	B'\left(p_0, p_{\perp}, p_z\right)&=& 
	\frac{1}{4}\frac{\left(-P.u\right)\text{Tr}(\Sigma(P)
		\slashed{P})+P^2\text{Tr}(\Sigma(P)\slashed{u})}{(P.u)^2-P^2},\\ \label{C}
	C'\left(p_0, p_{\perp}, p_z\right)&=& 
	-\frac{1}{4}\text{Tr}(\gamma_5\Sigma(P)\slashed{u}),\\ \label{D}
	D'\left(p_0, p_{\perp}, p_z\right)&=& 
	\frac{1}{4}\text{Tr}(\gamma_5\Sigma(P)\slashed{b}).
	\end{eqnarray}
	The explicit form of the above structure functions
	have been calculated as~\cite{Das:PRD97'2018}
	\begin{eqnarray}\label{A_net}
	A'(p_0,|{\bf{p}}|) &=& 
	\frac{m_{th}^2}{|{\bf{p}}|^2}Q_1\Bigg(\frac{p_0}{|{\bf{p}}|}\Bigg),\\
	B'(p_0,|{\bf{p}}|) &=&
	- \frac{m_{th}^2}{|{\bf{p}}|}\Bigg[\frac{p_0}
	{|{\bf{p}}|}Q_1\Big(\frac{p_0}{|{\bf{p}}|}\Big)
	-Q_0\Bigg(\frac{p_0}{|{\bf{p}}|}\Bigg)\Bigg],\\
	C'(p_0,|{\bf{p}}|) &=& -4g^2C_FM^2\frac{p_z}
	{|{\bf{p}}|^2}Q_1\Bigg(\frac{p_0}{|{\bf{p}}|}\Bigg),\\
	D'(p_0,|{\bf{p}}|) &=& -4g^2C_FM^2\frac{1}
	{|{\bf{p}}|}Q_0\Bigg(\frac{p_0}{|{\bf{p}}|}\Bigg),\label{D_net}
	\end{eqnarray}
	where
	\begin{eqnarray}
	M^2(T,\mu , B) = \frac{|q_iB|}{16\pi^2}
	\left(\frac{\pi T}{2m_{i0}} -\ln{2}  + 
	\frac{7\mu^2\zeta(3)}{8\pi^2T^2}\right),\label{magM}
	\end{eqnarray}
	where $\zeta$ corresponds to the Riemann zeta function.
	$Q_0$ and $Q_1$ refer to the  Legendre functions of first
	and second kind, respectively, which are given by
	\begin{eqnarray}
	Q_0(x) &=& \frac{1}{2}\ln\left(\frac{x+1}{x-1}\right),\\
	Q_1(x) &=& \frac{x}{2}\ln\left(\frac{x+1}{x-1}\right)-1 = xQ_0(x)-1.
	\end{eqnarray}
	The  quark self energy 
	can be  recast in the basis of right and left-hand chiral
	projection operators as 
	\begin{eqnarray}\label{63}
	\Sigma(P) = -P_R(A'\slashed{P} + 
	(B'+C')\slashed{u} + 
	D'\slashed{b})P_L - P_L(A'\slashed{P} + 
	(B'-\mathcal{C})\slashed{u} -
	D'\slashed{b})P_R.
	\end{eqnarray}
	We can re-write the inverse fermion propagator~\eqref{inverse_prop}
	using~\eqref{63} as
	\begin{eqnarray}
	S^{*-1}(P) = \slashed{P} + P_R\left[A'\slashed{P}
	+ \left(B'+C'\right)\slashed{u} + 
	D'\slashed{b}\right]P_L + P_L
	\left[A'\slashed{P} + 
	\left(B'-C'\right)\slashed{u} - 
	D'\slashed{b}\right]P_R,
	\end{eqnarray}
	which can further be simplified as
	\begin{eqnarray}
	S^{*-1}(P) = P_R \slashed{L} P_L + P_L \slashed{R} P_R.
	\label{inve_PR}		
	\end{eqnarray}
	Since $P_{L,R}\gamma^{\mu} = \gamma^{\mu} P_{R,L}$ and 
	$P_L \slashed{P} P_L = P_R \slashed{P} P_R = P_L P_R \slashed{P} = 0$, 
	$\slashed{L}$ and $\slashed{R}$ are given by
	\begin{eqnarray}
	\slashed{L} &=& (1+A')\slashed{P} +
	(B'+C')\slashed{u}
	+ D'\slashed{b},\\
	\slashed{R} &=& (1+A')\slashed{P} +
	(B'-C')\slashed{u} - D'\slashed{b}.
	\end{eqnarray}
	After inverting Eqn.~\eqref{inve_PR}, 
	we get the effective quark propagator as
	\begin{eqnarray}
	S^*(P) = \frac{1}{2}\left[ P_L\frac{\slashed{L}}
	{L^2/2}P_R+P_R\frac{\slashed{R}}{R^2/2}P_L\right],
	\end{eqnarray}
	where 
	\begin{eqnarray}
	& L^2 = (1+A')^2P^2 + 2(1+A')
	(B'+C')p_0-2D'(1
	+A')p_z+ (B'+C')^2-D'^2,\\
	& R^2 = (1+A')^2P^2 + 2(1+A')
	(B'-C')p_0+2D'
	(1+A')p_z+ (B'-C')^2-D'^2.
	\end{eqnarray}
	Now in order to get the quark thermal 
	mass in weakly magnetized thermal 
	QCD medium, we take the static limit
	($p_0 =0,|{\bf{p}}|\rightarrow 0$) of
	$L^2/2$ and $R^2/2$ modes,\footnote{We have expanded 
		the Legendre functions 
		appearing  in the structure functions in power series of 
		$\frac{|\bf{p}|}{p_0}$ and have  
		considered terms only up to $\mathcal{O}(g^2)$}
	we get (suppressing the flavor index)
	\begin{eqnarray}
	&\frac{L^2}{2}\arrowvert_{p_0= 0, {|\bf{p}|}\rightarrow 0}
	= m_{T}^2 + 4g^2 C_F M^2,\\
	&\frac{R^2}{2}|_{p_0= 0, {|\bf{p}|}\rightarrow 0}
	= m_{T}^2 - 4g^2 C_F M^2,
	\end{eqnarray}
where, $m_T$ and $M$ are as defined in Eqs.\eqref{quarkmassT} and \eqref{magM}, respectively. Here, we can infer that the otherwise degenerate left- and right- handed modes get separated out in presence of weak magnetic field as
	\begin{align}\label{qaurk_mass}
	&m_{L}^2 = m_{T}^2 + 4g^2 C_F M^2,\\
	&m_{R}^2 = m_{T}^2 - 4g^2 C_F M^2.
	\end{align}
	We will use these thermally generated masses in the 
	dispersion relation of the quarks to 
	calculate the Seebeck and Nernst coefficients in the 
	forthcoming sections.

\section{Thermoelectric response of an 
	anisotropic QCD medium}\label{three}
QGP produced in relativistic heavy ion collisions can possess a significant temperature gradient between its central and peripheral regions. A temperature-gradient and a finite chemical potential 
in a conducting medium create the necessary conditions for the Seebeck effect. Charge
carriers diffuse from regions of higher temperature to regions of lower temperature. This diffusion of charge carriers constitutes the Seebeck current, which leads to the generation of an
electric field. The diffusion ceases
when the strength of the created electric
field balances the thermodynamic gradient. The magnitude of electric field thus generated per unit temperature 
gradient in the medium is termed as the Seebeck coefficient and is evaluated 
in the limit of zero electric current\cite{Callen1960,Scheidemantel:PRB68'2003}. 
The Seebeck coefficient is a quantitative estimate of the 
efficiency of conversion of a temperature gradient into electric field 
by a conducting medium. 
The sign of the Seebeck coefficient can be used to determine the sign 
of majority charge carriers in condensed matter systems, as it is 
positive for positive charge carriers and negative for negative charge 
carriers.  Upcoming experimental programs such as the Facility for Antiproton 
and Ion Research (FAIR) in Germany and the  Nuclotron-based Ion Collider fAcility 
(NICA) in Russia, where low-energy heavy ion collisions are expected to create a 
baryon-rich plasma, could be the perfect environment for the aforementioned 
thermoelectric phenomenon to manifest.

In the presence of a magnetic field, the charged particles drift perpendicular to their original direction of motion due to the Lorentz force acting on them. This leads to a thermocurrent that is transverse to both the direction of temperature gradient and the external magnetic field. This is called the Nernst effect.  Like the Seebeck coefficient, the Nernst coefficient is also calculated at the limit of zero electric current, that is, by enforcing the equilibrium condition. The Nernst coefficient can be defined as the electric field induced in the $\hat{x}$ ($\hat{y}$) direction per unit temperature gradient in the $\hat{y}$ ($\hat{x}$) direction, in the presence of a magnetic field pointing in the $\hat{z}$ direction.

 Due to a larger expansion rate of the medium along the longitudinal direction
	compared to the radial direction, one  develops a local momentum anisotropy. This anisotropy can be taken into account by introducing an anisotropy parameter $\xi$ in the isotropic distribution function. This is the Romatschke-Strickland (RS) parametrisation of the anisotropic distribution function with a single anisotropy parameter. For the case of weak momentum anisotropy
	($\xi<1$), and a finite quark chemical potential $\mu$, the form of the RS distribution is given as (suppressing the flavor index) \cite{Romatschke:PRD68'2003}. 
	
	\ba\label{A.D.F.}
	f_a^0(\mathbf{p};T)=\frac{1}{e^{\beta\left(\sqrt{\rm{p}^2+\xi(\mathbf{p}\cdot\mathbf{n})^2+m^2}\,\,-\mu\right)}+1}
	~,\ea
which can be expanded in a Taylor series about $\xi=0$ (isotropic case). Keeping terms upto linear power in $\xi$, we have:
	\begin{eqnarray}\label{fa0}
	f_a^0=f^0- \xi \beta \frac{\bf{(p.n)}^2}{2\epsilon}f^0(1-f^0),
	\end{eqnarray}
	with 
	\begin{equation}
	f^0=f_a^0(\xi=0)=\frac{1}{e^{\beta\left(\sqrt{\rm{p}^2+m^2}\,-\mu\right)}+1},
	\end{equation}
where, $\beta=1/T$.	The anisotropy parameter $\xi$ is defined as
	\begin{eqnarray}
	\xi=\frac{<p_{T}^2>}{2<p_{L}^2>}-1,
	\end{eqnarray}
	where, $p_L$ and $\bm{p_T}$ refer to the longitudinal and transverse components of $\bm{p}$. The 2 in the denominator denotes the fact that there are two transverse directions with respect to any given vector. The condition of isotropy is when $<p_{T}^2>=2<p_{L}^2>$. For $p_T> 2p_L$, $\xi$ is positive. The aforementioned components are defined with respect to an arbitrary anisotropy direction, denoted by the vector $\bm{n}=(\sin{\alpha},0,\cos{\alpha})$, with $\alpha$ being the angle between the direction of anisotropy and the $z$-axis. This parameter is arbitrary and hence physical quantities should be independent of it. The longitudinal and transverse momentum components are then defined as $p_{L}={\bf p.n}, {\bf p_{T}=p-  n.(p.n)}$. In spherical polar coordinates, ${\bf p}=(p\sin{\theta}\cos{\phi},p\sin{\theta}\sin{\phi},
	p\cos{\theta} )$, where, $\theta$ and $\phi$ are the polar and azimuthal angles, respectively.  
	${\bf (p.n)^2}= p^2 c(\theta, \alpha, \phi)= p^2 
	(\sin^2{\alpha} \sin^2{\theta} \cos^2{\phi}+
	\cos^2{\alpha} \cos^2{\theta} +\sin{2\alpha} \sin{\theta}
	\cos{\theta}\cos{\phi})$.  Thus, Eq.\eqref{fa0} ultimately becomes
	\begin{equation}\label{fa0_f}
	f_a^0=f^0- \xi \beta \frac{p^2c(\theta,\alpha,\phi)}{2\epsilon}f^0(1-f^0).	
	\end{equation}
\subsection{In the absence of magnetic field}
In this section, we evaluate the thermoelectric response of the anisotropic QGP medium in the absence of a background magnetic field within the kinetic theory framework.  Our starting point is
the evolution of the single particle distribution 
function, which is given by the relativistic Boltzmann 
transport equation (RBTE). In the relaxation time approximation (RTA), the equation reads

\begin{eqnarray}
p^{\mu}\frac{\partial f}{\partial x^{\mu}}+q~F^{\rho \sigma}
p_{\sigma}\frac{\partial f}{\partial p^{\rho}}= 
-\frac{p^{\mu}u_{\mu}}{\tau}\left(f_a-f^0_a \right),
\label{rbte}
\end{eqnarray}
where, $f_a^0$ is as defined in Eq.\eqref{fa0}. In the absence of a magnetic field, we only have the $\rho = i$, $\sigma = 0$ and $\sigma=1$, $\rho=0$ components of the electromagnetic field strength tensor $F^{\rho\sigma}$. These components are $F^{0i}=-\bm{E}$, $F^{i0}=\bm{E}$. The RBTE [Eq.\eqref{rbte}] then takes the form
\begin{eqnarray}\label{rbte2}
p^0. \frac{\partial f_a}{\partial t}
+ \bm{p}.\frac{\partial f_a}{\partial \bm{r}}+
q{\bf E.p} \frac{\partial f_a}{\partial p^0}+
qp_0{\bf E}. \frac{\partial f_a}{\partial{\bf p}}&=&
-\frac{p^{\mu}u_{\mu}}{\tau}\left(f_a-f^0_a \right).
\label{rbte_b0}
\end{eqnarray}
Considering the particles to be on-shell, we use the chain rule of differentiation 
\[	\frac{\partial}{\partial \bm{p}}\rightarrow \frac{\partial p^0}{\partial \bm{p}}\frac{\partial}{\partial p^0}+\frac{\partial}{\partial \bm{p}}=\frac{\bm{p}}{p^0}\frac{\partial}{\partial p^0}+\frac{\partial}{\partial \bm{p}},\]
by which, the RBTE in the local rest frame becomes :
\begin{equation}
\frac{\partial f_a}{\partial t}+	\bm{p}.\frac{\partial f_a}{\partial \bm{r}}+
	q{\bf E}\cdot \frac{\partial f_a}{\partial \bm{p}}=-\frac{\delta f_a}{\tau}\label{rbte3}
\end{equation}

where $f_a=f^0_a+\delta f_a$, $\delta f_a$ being the 
deviation of the distribution function from the ``anisotropic equilibrium"\cite{Bazow:PRC90'2014} distribution $f_a^0$ (defined in Eq.\ref{A.D.F.}), with $m$ as defined in Eq.\eqref{para_massT}. Thus, the distribution function is expanded in an anisotropic background, wherein the leading order term ($f_a^0$) itself is dissipative. The form of $f_a^0$ used here was first proposed by Romatschke and Strickland, and has a spheroidal symmetry. $\delta f_a$ is a measure of the departure of the distribution function from this symmetry.
Further, in this study, we have considered the system to be
close to equilibrium, hence we have taken
  small values of the 
anisotropy parameter
 $(\xi = 0.3,0.6)$. 
   
In addition, we are interested in the steady state so 
the first term in Eqn.~\eqref{rbte3} vanishes, finally leaving us with
\begin{equation}
	\bm{p}.\frac{\partial f_a}{\partial \bm{r}}+
	q{\bf E}\cdot \frac{\partial f_a}{\partial \bm{p}}=-\frac{\delta f_a}{\tau}\label{rbte4}
\end{equation}

  $\tau$ is  the relaxation time. Once the system is infinitesimally disturbed from equilibrium, it takes an average an amount of time $\tau$ to revert to equilibrium.
The relaxation time has been   
calculated for the quarks considering $2\to 2$ scatterings using  perturbative QCD as~\cite{Hosoya:NPB250'1985}
\ba
\tau(T) =\frac{1}{5.1T \alpha_s^2 \log \left(\frac{1}{\alpha_s}\right)
	\label{tau_B0} 
	[1+0.12(2N_f+1)]},  
\ea 
where $\alpha_s$ is the running coupling 
constant~\eqref{coupling_T}.\par
In order to calculate the deviation $\delta f$, we 
assume that the system deviates only infinitesimally away from equilibrium, \textit{i.e.} $ \delta f \ll f^0$.
We then compute the relevant derivatives required to evaluate the left hand side of 
Eq.~\eqref{rbte4}:
\begin{eqnarray}
\frac{\partial f_a^0}{\partial \bm{p}}&=&\frac{\partial f^0}{\partial \bm{p}}
\left[1-\xi \beta \frac{c}{2}\left \{ 
\frac{p^2}{\beta \epsilon^2}-\frac{2}{\beta}+
\frac{p^2}{\epsilon}-\frac{2p^2}{\epsilon}f^0\right\} \right]
=\frac{\partial f^0}{\partial \bm{p}}L_1(p,\xi)\\
\label{dev1}
\frac{\partial f_a^0}{\partial \epsilon}&=&\frac{\partial f^0}
{\partial \epsilon}
\left[1-\xi \beta \frac{c}{2}\left \{ 
\frac{p^2}{ \epsilon}-\frac{2f^0p^2}{\epsilon}+
\frac{p^2}{\beta \epsilon^2}\right\} \right]
=\frac{\partial f^0}{\partial \epsilon}L_2(p,\xi)\\
\frac{\partial f_a^0}{\partial \bm{r}}&=&\frac{\partial f^0}{\partial \bm{r}}
\left[1-\xi \beta \frac{c}{2}\left \{ 
\frac{p^2}{\epsilon}-\frac{2 f^0p^2}{\epsilon}-
\frac{p^2}{\epsilon(\epsilon-\mu)}\right\} \right]
=\frac{\partial f^0}{\partial \bm{r}}L_3(p,\xi),
\end{eqnarray}
where, $\epsilon=\sqrt{p^2+m^2}$. As expected, the above derivatives reduce to their isotropic expressions on putting $\xi=0$.  In the first approximation, $f_a$ is replaced with $f_a^0$ in the LHS of Eq.\eqref{rbte4}. We then substitute the above derivatives in 
Eq.~\eqref{rbte4} to get $\delta f$
\ba 
\delta f_a =-\beta^2 \tau (\epsilon-\mu)
f_0(1-f_0)~{\bf \frac{p}{\epsilon}.\nabla T } L_3(p,\xi)
+\tau q \frac {\bf E.p}{\epsilon}\beta f_0(1-f_0)L_1(p,\xi).
\ea
Similarly, we calculate the deviation
$\delta \bar{f}$ for the anti quarks. This is done by replacing the quark distribution function $f$ by the anti-quark distribution function $\bar{f}$ and changing the sign of the chemical potential $\mu$. 
\ba 
\delta \bar{f_a} =-\beta^2 \tau (\epsilon +\mu)
\bar{f}_0(1-\bar{f}_0) L_3(p,\xi)~{\bf \frac{p}{\epsilon}.\nabla T }
+\tau \bar{q} \frac {\bf E.p}{\epsilon}L_1(p,\xi)\beta \bar{f}_0(1-\bar{f}_0),
\ea
where, the anti-quark isotropic distribution function is 
\begin{equation}
\bar{f^0}=\frac{1}{e^{\beta(\epsilon+\mu)+1}}.
\end{equation}
We consider the temperature gradient to exist in the $x$-$y$ plane, \textit{i.e.}
\begin{equation}
	\bm{\nabla}T=\frac{\partial T}{\partial x}\hat{\bm{x}}+\frac{\partial T}{\partial x}\hat{\bm{y}.}
\end{equation}
Consequently, the induced electric field is also considered to be planar, \textit{i.e.}, $\bm{E}=E_x\hat{\bm{x}}+E_y\hat{\bm{y}}$. The induced four current due to a single quark
flavour can be written as 
\begin{eqnarray}
J^{\mu}=g \int \frac{d^3p}{(2\pi)^3}\frac{p^{\mu}}{\epsilon}
[q\,\delta f_a+\bar{q}\,\delta \bar{f_a}],
\label{current_weak}
\end{eqnarray}
where, $g$ is the quark degeneracy factor.

Now substituting $\delta f$ and $\delta \bar{f}$ in Eq.\eqref{current_weak} and putting the induced current to zero in the steady state we get
\ba\label{sc}
\bm{E}=S \bm{\nabla} T. 
\ea
Here, $S$ is the \textit{individual} Seebeck coefficient, \textit{i.e.} the Seebeck coefficient of a hypothetical medium consisting of a single quark flavor. 
\ba 
S=-\frac{H_1}{H_2}
\ea
where
\begin{align}
H_{1}&=\frac{qg\beta^2}{3}\int \frac{d^3p}{(2\pi)^3}~\frac{p^2\tau}{\epsilon^2}
\bigg\{
(\epsilon +\mu)\bar{f}_0(1-\bar{f}_0)L_3(p,\xi)-(\epsilon -\mu)f_0(1-f_0)L_3(p,\xi)\bigg \}\\[0.5em]
H_{2}&=\frac{q^2g\beta}{3} \int \frac{d^3p}{(2\pi)^3}~\frac{p^2\tau}{\epsilon^2} \bigg \{
f_0(1-f_0)L_1(p,\xi)+\bar{f}_0(1-\bar{f}_0)L_1(p,\xi) \bigg \}.
\end{align}
To evaluate the Seebeck coefficient of the composite medium, we need to take into account the total current due to multiple quark species. The spatial part of the total 4-current is given by
\begin{equation}
	 \bm{J}=\sum_iq_ig_i\int \frac{d^3\mbox{p}}{(2\pi)^3\epsilon}\bm{p}\left[\delta f_a^i-\overline{\delta f_a^i}\right]\label{totalJ}
\end{equation}
Setting the $x$ and $y$ components of the above equation to zero yields the following equations:
\ba \label{Jxeq} 
\sum_{i=u,d}\left[(H_2)_i E_x+(H_1)_i\frac{\partial T}{\partial x}\right]
&=&0, \\
\sum_{i=u,d}\left[(H_2)_iE_y
+(H_1)_i\frac{\partial T}{\partial y}\right]&=&0, 
\label{Jyeq}
\ea
Solving the above equations and comparing with Eq.\eqref{sc}, we get
\begin{equation}
\begin{pmatrix}
	E_x \\
	E_y
\end{pmatrix}
=\begin{pmatrix}
	S_{\text{tot}} & 0\\
	0 & S_{\text{tot}}
\end{pmatrix}
\begin{pmatrix}
	\frac{\partial T}{\partial x}~\\
	\frac{\partial T}{\partial y},
\end{pmatrix}\label{matrix}
\end{equation}

where, the total Seebeck coefficient $S_{tot}$ is given by
\begin{equation}
S_{tot} = -\frac{C_1C_2}{C_2^2},
\end{equation}
with
\begin{equation}
	C_1=\sum_{i=u,d} (H_1)_i,  \quad \quad 
	C_2=\sum_{i=u,d} (H_2)_i
\end{equation}
In the next subsection, we will see how the weak magnetic field
in the background modulates the thermoelectric response of the 
hot QCD medium.
\subsection{In the presence of weak magnetic field}
The RBTE [Eq.\eqref{rbte}] in 3-vector notation, in the presence of the 
Lorentz force can be written as
\begin{eqnarray}
\frac{\partial f_a}{\partial t}+\bm{v}.
\frac{\partial f_a}{\partial \bm{r}}+
q~(\bm{E}+\bm{v}\times \bm{B})\cdot \frac{\partial f_a}{\partial \bm{p}}&=&
-\frac{\delta f_a}{\tau},
\label{rbte_weak}
\end{eqnarray}
In the first approximation, we replace $f_a$ by $f_0$ in the L.H.S. above, similar to the $B=0$ case.   We know that 
the derivative
$\frac{\partial f^0}{\partial \bm{p}} \varpropto \bf{v}$. 
The term  $(\bf{v} \times \bf{B}). \frac{\partial f_a^0}{\partial \bm{p}}$ in Eqn.~\eqref{rbte_weak} can be replaced with $(\bf{v} \times \bf{B}). 
\frac{\partial f^0}{\partial p}L_1(p,\xi)$ according to 
Eqn.~\eqref{dev1}. Thus, $\frac{\partial f_a^0}{\partial \bm{p}}$ is also proportional to $\bm{v}$, and therefore, the term $(\bf{v} \times \bf{B}). \frac{\partial f_a^0}{\partial \bm{p}}$ also vanishes. 
 The contribution to the Lorentz force thus comes solely from $\delta f_a$. Additionally, we work in the static approximation where, both $f_a$ and $f_a^0$ do not depend on time, so that the first term in Eq.\eqref{rbte_weak} drops out. The RBTE thus gets reduced to
\begin{equation}\label{3rbte}
	\bm{v}\cdot
	\frac{\partial f^0_a}{\partial \bm{r}}+
	q~\bm{E}\cdot \frac{\partial f^0_a}{\partial \bm{p}}+q(\bm{v}\times \bm{B})\frac{\partial (\delta f_a)}{\partial \bm{p}}=
	-\frac{\delta f_a}{\tau}
\end{equation}
 As earlier, $f_a=f_a^0+\delta f_a$ with $\delta f_a\ll f_a^0.$ To solve for $\delta f_a$ we take an ansatz, similar to a trial solution for solving any differential equation\cite{Das:PRD102'2020}:
\begin{eqnarray}\label{ansatz}
\delta f_a=({\bf p .\Sigma})~\frac{\partial f^0_a}{\partial \epsilon}
\end{eqnarray}
 where $ \bm{\Sigma}$ depends on the temperature gradient, magnetic field and electric field. The general form of $ \bm{\Sigma}$ can be written as 
\begin{eqnarray}\label{sigma}
{\bf \Sigma} =\alpha_1 {\bf E}+\alpha_2 {\bf b}+\alpha_3
{\bf (E\times b)}+
\alpha_4 {\bf \nabla T}+\alpha_5 {\bf (\nabla T \times b)}+
\alpha_6 {\bf (\nabla T \times E)},
\end{eqnarray}
which is basically an expansion in terms of all possible vector formations from those available in the system. Here, $\alpha_1$, $\alpha_2$, $\alpha_3$, $\alpha_4$, $\alpha_5$
and $\alpha_6$ are unknown coefficients which need to be solved for. Using Eqs.\eqref{sigma} and \eqref{ansatz} in Eq.\eqref{3rbte}, we get,

\ba
\beta^2 (\epsilon-\mu)f_0(1-f_0) L_3(p,\xi)~{\bf v.\nabla T }
-\beta q f_0(1-f_0) L_1(p,\xi) ~{\bf v.E}
-\beta qf_0(1-f_0) L_2(p,\xi) \nonumber\\
\{-\alpha_1 |B|{\bf v.(E \times b)}
+\alpha_3 |B|~{ \bf v. E }-
\alpha_4 |B| {\bf v. (\nabla T \times b)}+
\alpha_5 |B| ~{\bf v. \nabla T }\} = \nonumber\\
f^0(1-f^0) L_2(p,\xi)\frac{\epsilon}{\tau}\left\{\alpha_1 {\bf v.E}
+\alpha_2 {\bf v.b}+\alpha_3
{\bf v.(E\times b)}+
\alpha_4 {\bf v.\nabla T}+\alpha_5 {\bf v.(\nabla T \times b)}
\right \}
\ea
Now, comparing the coefficients
of tensor structures from both sides of the above equation, we get
\ba
\frac{\epsilon}{\tau}\alpha_1 L_2(p,\xi) &=& 
-\alpha_3 q|B|L_2(p,\xi)-qL_1(p,\xi)  \\
\frac{\epsilon}{\tau}\alpha_2 L_2(p,\xi) &=& 0\\
\frac{\epsilon}{\tau}\alpha_3 L_2(p,\xi)&=&
\alpha_1 q|B|L_2(p,\xi) \\
\frac{\epsilon}{\tau}\alpha_4 L_2(p,\xi) &=&  
\beta (\epsilon -\mu)L_3(p,\xi)
-\alpha_5 q|B|L_2(p,\xi) \\
\frac{\epsilon}{\tau}\alpha_5 L_2(p,\xi) &=& \alpha_4 q|B|L_2(p,\xi)
\ea

We find the values of $\alpha_1$, $\alpha_2$, 
$\alpha_3$, $\alpha_4$ and  $\alpha_5$ from the 
above equations, which come out to be

\ba
\alpha_1 &=& -\frac{\tau}{\epsilon}\frac{q}
{(1+\omega_c^2 \tau^2)}\frac{L_1(p,\xi)}{L_2(p,\xi)}\\
\alpha_2 &=& 0\\
\alpha_3 &=& -\frac{\tau^2}{\epsilon}\frac{\omega_c q}
{(1+\omega_c^2 \tau^2)}\frac{L_1(p,\xi)}{L_2(p,\xi)}\\
\alpha_4 &=& \frac{\tau}{\epsilon}\frac{\beta (\epsilon-\mu)}
{(1+\omega_c^2 \tau^2)}\frac{L_3(p,\xi)}{L_2(p,\xi)} \\
\alpha_5 &=& \frac{\tau^2}{\epsilon}\frac{\omega_c \beta (\epsilon-\mu)}
{(1+\omega_c^2 \tau^2)}\frac{L_3(p,\xi)}{L_2(p,\xi)}
\ea
\ba
\delta f &=& \frac{\tau}{(1+\omega_c^2 \tau^2)} {\bf p.}
\bigg \{\left(-\frac{q}{\epsilon}
{\bf E}-\frac{q\tau \omega_c}{\epsilon}{\bf (E\times b)}\right)\frac{L_1(p,\xi)}{L_2(p,\xi)}\\
&+& \left(\frac{\beta (\epsilon -\mu)}{\epsilon}
{\bf \nabla T}+\frac{\beta \tau \omega_c (\epsilon -\mu)}{\epsilon}
{\bf (\nabla T \times b)}\right)\frac{L_3(p,\xi)}{L_2(p,\xi)}\bigg \} 
\frac{\partial f^0}{\partial \epsilon}L_2(p,\xi)
\ea  
Similarly,we can compute the deviation in the anti quarks
distribution function as
\ba
\delta \bar{f} &=& \frac{\tau}{(1+\omega_c^2 \tau^2)} {\bf p.}
\bigg \{\left(\frac{q}{\epsilon}
{\bf E}-\frac{q\tau \omega_c}{\epsilon}
{\bf (E\times b)}\right)
\frac{\bar{L}_1(p,\xi)}{\bar{L}_2(p,\xi)}\\
&+& \left(\frac{\beta (\epsilon +\mu)}{\epsilon}
{\bf \nabla T}-\frac{\beta \tau \omega_c (\epsilon +\mu)}{\epsilon}
{\bf (\nabla T \times b)}\right)
\frac{\bar{L}_3(p,\xi)}{\bar{L}_2(p,\xi)}\bigg \} 
\frac{\partial \bar{f}^0}{\partial \epsilon}L_2(p,\xi)
\ea
The $x$ and $y$ components of the induced current density
can be written as 
\ba
J_x &=&I_1 E_x+I_2E_y+I_3\frac{\partial T}{\partial x}
+I_4\frac{\partial T}{\partial y}\label{J_x}\\
J_y &=& -I_2 E_x+I_1E_y-I_4\frac{\partial T}{\partial x}
+I_3\frac{\partial T}{\partial y} \label{J_y}
\ea 

The electric 
field components are related to the temperature
gradients, Seebeck and Nernst coefficients
via a matrix equation 
\ba
\begin{pmatrix}
	E_x \\
	E_y
\end{pmatrix}
=\begin{pmatrix}
	S & N|B|\\
	-N|B| & S
\end{pmatrix}
\begin{pmatrix}
	\frac{\partial T}{\partial x}~\\
	\frac{\partial T}{\partial y}
\end{pmatrix}\label{nernst}
\ea
where
\begin{align}
I_{1}&=\frac{q^2g\beta}{3} \int \frac{d^3p}{(2\pi)^3}~\frac{p^2}{\epsilon^2}
\frac{ \tau}{(1+\omega_c^2 \tau^2)} \bigg \{
f_0(1-f_0)L_1(p,\xi)+\bar{f}_0(1-\bar{f}_0)L_1(p,\xi) \bigg \}\\
I_{2}&=\frac{q^2g\beta}{3}\int \frac{d^3p}{(2\pi)^3}~\frac{p^2}{\epsilon^2}
\frac{\omega_c \tau^2}{(1+\omega_c^2 \tau^2)}\bigg \{
f_0(1-f_0)L_1(p,\xi)-\bar{f}_0(1-\bar{f}_0)L_1(p,\xi)\bigg \}\\
I_{3}&=\frac{qg\beta^2}{3}\int \frac{d^3p}{(2\pi)^3}~\frac{p^2}{\epsilon^2}
\frac{\tau}{(1+\omega_c^2 \tau^2)}\bigg\{
(\epsilon +\mu)\bar{f}_0(1-\bar{f}_0)L_3(p,\xi)\nonumber\\&-(\epsilon -\mu)f_0(1-f_0)L_3(p,\xi)\bigg \}\\
I_{4}&=\frac{qg\beta^2}{3}\int \frac{d^3p}{(2\pi)^3}~\frac{p^2}{\epsilon^2}
\frac{\omega_c \tau^2}{(1+\omega_c^2 \tau^2)}\bigg \{-(\epsilon -\mu)
f_0(1-f_0)L_3(p,\xi)\nonumber\\&-(\epsilon +\mu)
\bar{f}_0(1-\bar{f}_0)L_3(p,\xi)\bigg \}
\end{align}

In the state of equilibrium, the components of the 
induced current density along $x$ and $y$ direction
vanish {\em i.e.} $J_x=J_y=0$. We can write 
from Eqns.~\eqref{J_x} and~\eqref{J_y}
\ba \label{J_x0} 
I_1 E_x+I_2E_y+I_3\frac{\partial T}{\partial x}
+I_4\frac{\partial T}{\partial y}&=&0, \\
-I_2 E_x+I_1E_y-I_4\frac{\partial T}{\partial x}
+I_3\frac{\partial T}{\partial y}&=&0, 
\label{J_y0}
\ea
We can further write Eqns~\eqref{J_x0} and \eqref{J_y0}  
as 
\ba
E_x=\left(-\frac{I_1I_3+I_2I_4}{I_1^2+I_2^2}\right)
\frac{\partial T}{\partial x} 
+\left(-\frac{I_2I_3-1_1I_4}{I_1^2+I_2^2}\right)
\frac{\partial T}{\partial y},\\
E_y=\left(-\frac{I_1I_3+I_2I_4}{I_1^2+I_2^2}\right)
\frac{\partial T}{\partial y} 
-\left(-\frac{I_2I_3-I_1I_4}{I_1^2+I_2^2}\right)
\frac{\partial T}{\partial x},
\ea
where,
\ba
S &=& -\frac{(I_1I_3+I_2I_4)}{I_1^2+I_2^2},\\[0.3em]
N|B|&=&\frac{(I_2I_3-I_1I_4)}{I_1^2+I_2^2}.
\ea
The integrals $I_2$ and $I_4$ vanishes in the 
absence of the magnetic field. As a result, the Nernst 
coefficient also vanishes.\\

In what follows, we will compute the Seebeck and 
Nernst coefficients for the medium composed of the $u$ and $d$ quarks. In the medium 
The 
$x$ and $y$ components of the induced current in the medium can then be written as the sum of the individual flavour contributions as
\vspace{-2mm}
\begin{eqnarray}
J_x&=&\sum_{i=u,d}\left[ (I_1)_iE_x+ (I_2)_iE_y
+ (I_3)_i\frac{\partial T}{\partial x}+
(I_4)_i\frac{\partial T}{\partial y}\right],\\
J_y&=&\sum_{a=u,d}\left[- (I_2)_iE_x+ (I_1)_iE_y-
(I_4)_i\frac{\partial T}{\partial x}+
(I_3)_i\frac{\partial T}{\partial y}
\right].
\end{eqnarray}
We extract the 
Seebeck and Nernst coefficients for the QCD medium
composed of u and d quarks by imposing the 
equilibrium condition ({\em i.e.} putting $J_x=J_y=0$) as
\ba
S^{B'}_{tot} &=& -\frac{(K_1K_3+K_2K_4)}{K_1^2+K_2^2},\\
N|B|&=&\frac{(K_2K_3-K_1K_4)}{K_1^2+K_2^2}.
\ea

\vspace{-2mm}
where,

\vspace{-9mm}
\ba
K_1=\sum_{i=u,d} (I_1)_i,  \quad \quad 
K_2=\sum_{i=u,d} (I_2)_i, \nonumber\\
K_3=\sum_{i=u,d}(I_3)_i,  \quad \quad 
K_4=\sum_{i=u,d} (I_4)_i.
\ea
\section{Results and discussion}
\vspace{-5mm}
We begin with the results obtained at $B=0$, followed by those obtained at finite $B$.
\begin{figure}
	\begin{subfigure}{0.48\textwidth}
		\includegraphics[width=0.95\textwidth,height=0.28\textheight]{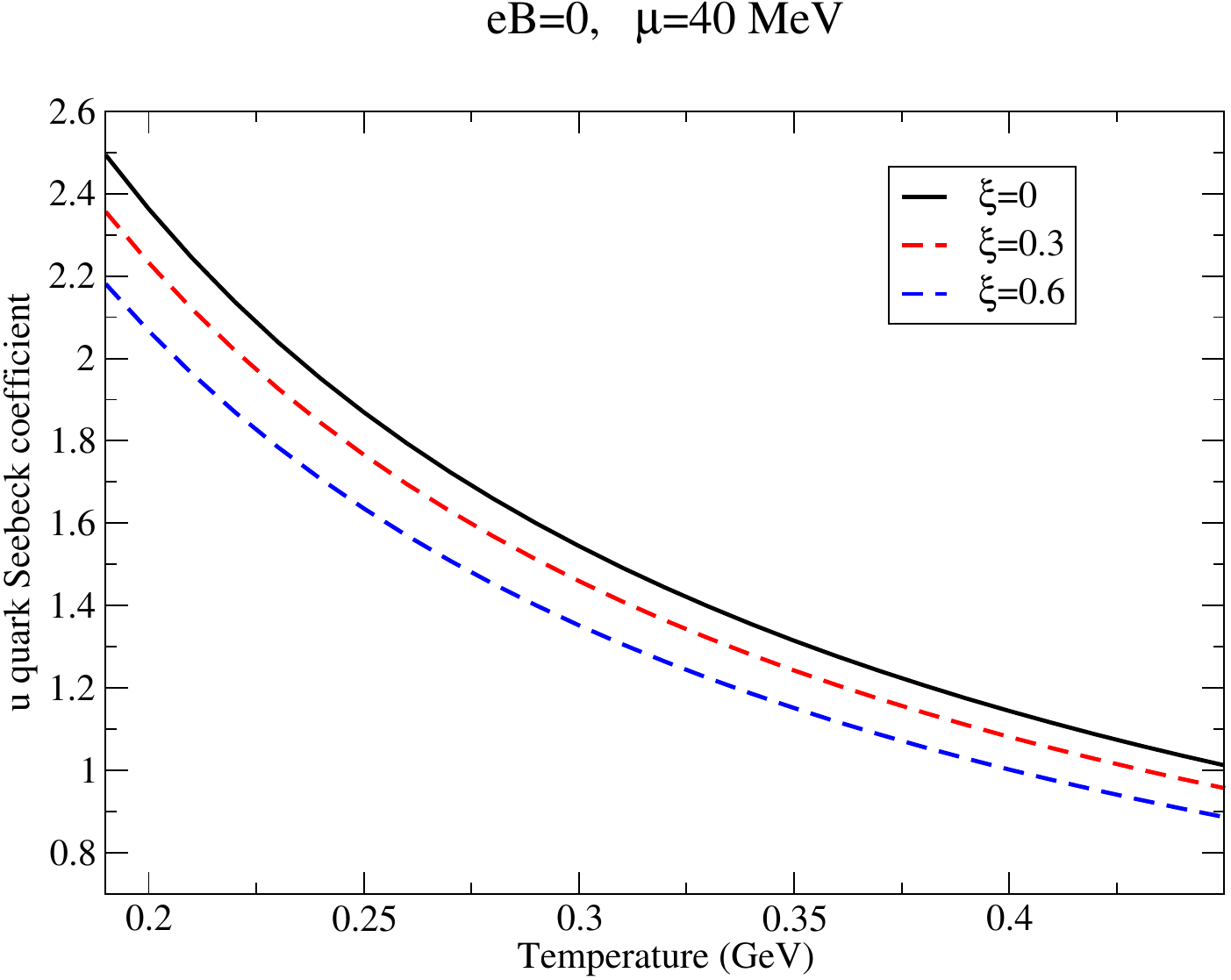}
		\caption{}\label{uSeebeckB0}
	\end{subfigure}
	\hspace*{\fill}
	\begin{subfigure}{0.48\textwidth}
		\includegraphics[width=0.95\textwidth,height=0.28\textheight]{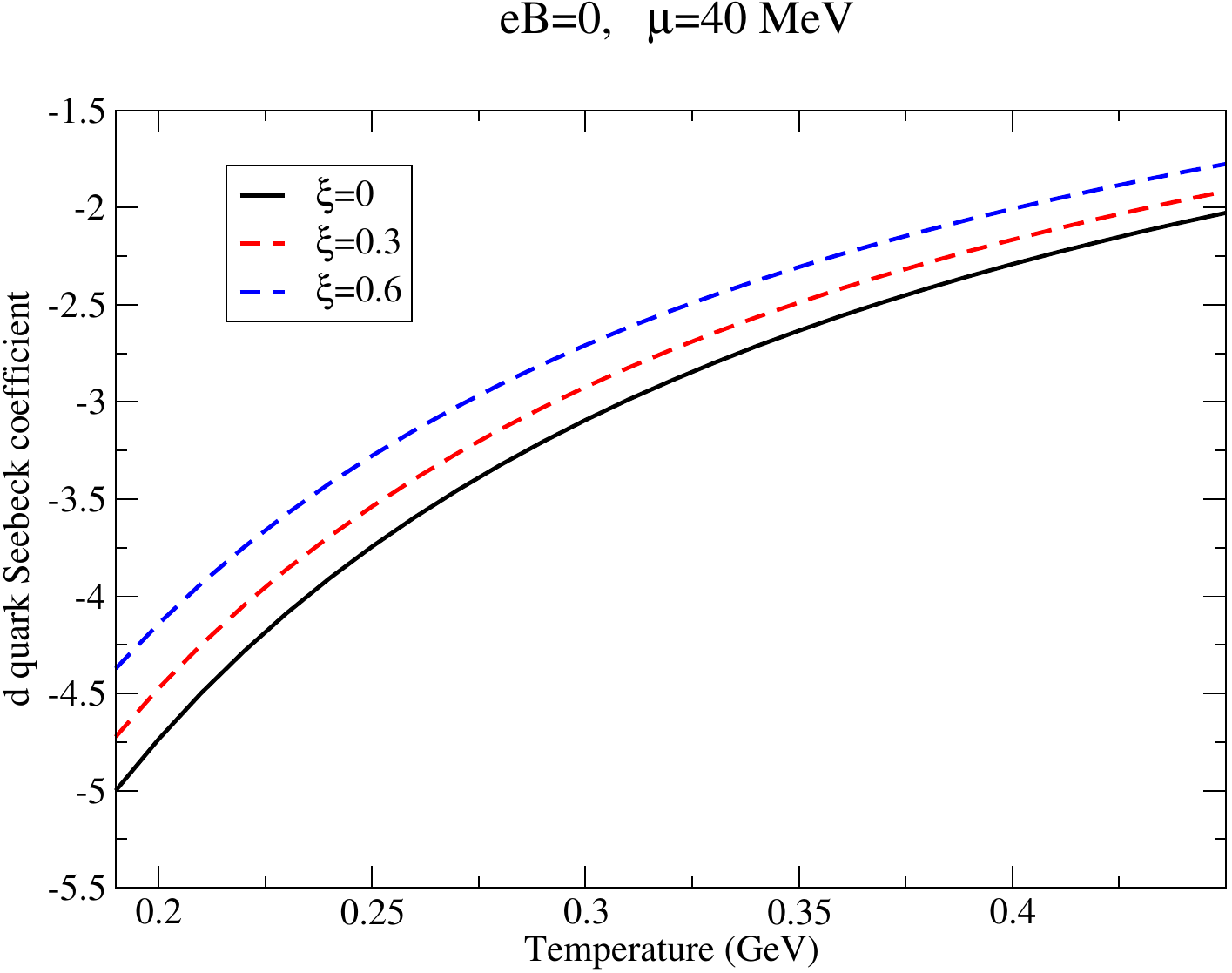}
		\caption{}\label{dSeebeckB0}
	\end{subfigure}
	\caption{(a) Temperature dependence of $u$ quark Seebeck coefficient in the absence of $B$, at a fixed value of $\mu$. (b) Temperature dependence of $d$ quark Seebeck coefficient in the absence of $B$, at a fixed value of $\mu$. The different curves correspond to different values of $\xi$.   }
	\label{ind_seebeckB0} 
\end{figure}
Figs. \eqref{uSeebeckB0} and \eqref{dSeebeckB0} show the temperature variation of individual Seebeck coefficients. That is to say, if the medium were composed exclusively of a single species of quark ($u$ or $d$), the Seebeck coefficients would vary with temperature as shown in the aforementioned figures. The positively charged $u$ quark gives rise to a positive Seebeck coefficient, while the same is negative in case of the negatively charged $d$ quark, which concurs with previous results\cite{Bhatt:PRD99'2018,Dey:PRD102'2020}. Positivity of the Seebeck coefficient indicates that the induced electric field is along the direction of the temperature gradient, whereas a negative value indicates that the induced electric field is in the direction opposite to the direction of the temperature gradient. It should be noted that we have used the convention where the direction of increasing temperature is considered positive. The magnitudes of both the individual Seebeck coefficients ($S_u$, $S_d$) decrease with temperature. Importantly, the magnitudes also decrease with the strength of anisotropy parameter $\xi$. $S_u$ decreases by 5.32\% while going from $\xi=0$ to $\xi=0.3$, averaged over the entire temperature range. From $\xi=0.3$ to $\xi=0.6$, the decrease is 7.11\%. Interestingly, the corresponding values for $S_d$ are identical up to two decimal places.
\begin{figure}
	\centering
	\includegraphics[width=7.5cm,height=7.5cm]{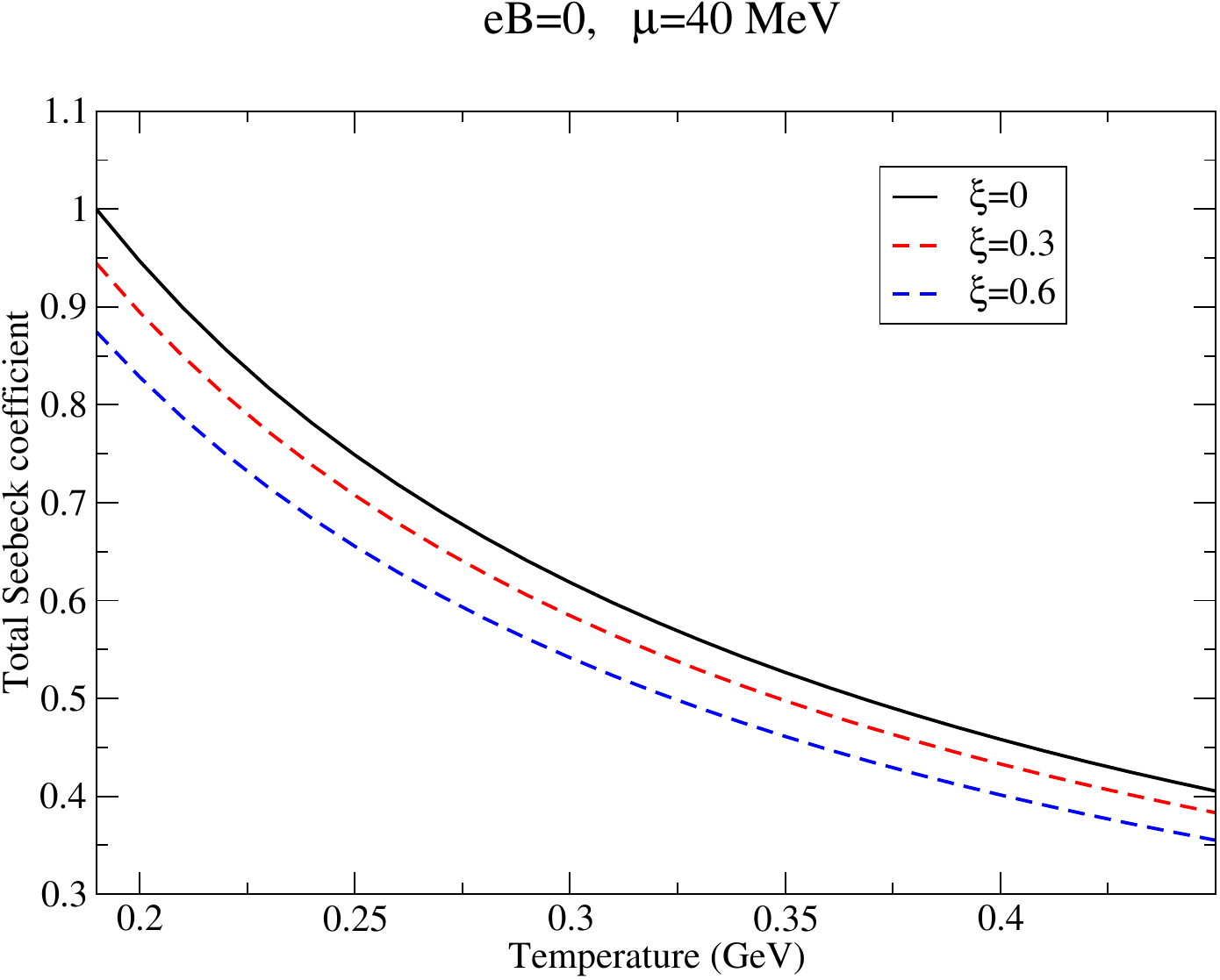}
	\caption{Seebeck coefficient of the composite medium in the absence of $B$ as a function of temperature.}
	\label{seebeckB0}
\end{figure}
Fig.\eqref{seebeckB0} shows the Seebeck coefficient of the composite medium composed of $u$ and $d$ quarks-the total Seebeck coefficient ($S_{\text{tot}}$), as a function of temperature. $S_{\text{tot}}$ is positive and decreases with temperature. So, the induced electric field points in along the temperature gradient. Also, it is to be noted that even though the magnitude of $d$ quark Seebeck coefficient is larger than that of the $u$ quark at a given temperature, the total coefficient is positive. This reflects the fact that a $u$ quark carries double the electric charge than a $d$ quark. Again, a finite anisotropy decreases the magnitude of $S_{\text{tot}}$. It decreases by 5.31\% in going from $\xi=0$ to $\xi=0.3$, and by 7.11\% when $\xi$ changes from 0.3 to 0.6. The decrease in the magnitude of transport coefficients with anisotropy has been observed earlier\cite{Rath:PRD100'2019}.

\begin{figure}
	\begin{subfigure}{0.48\textwidth}
		\includegraphics[width=0.95\textwidth,height=0.3\textheight]{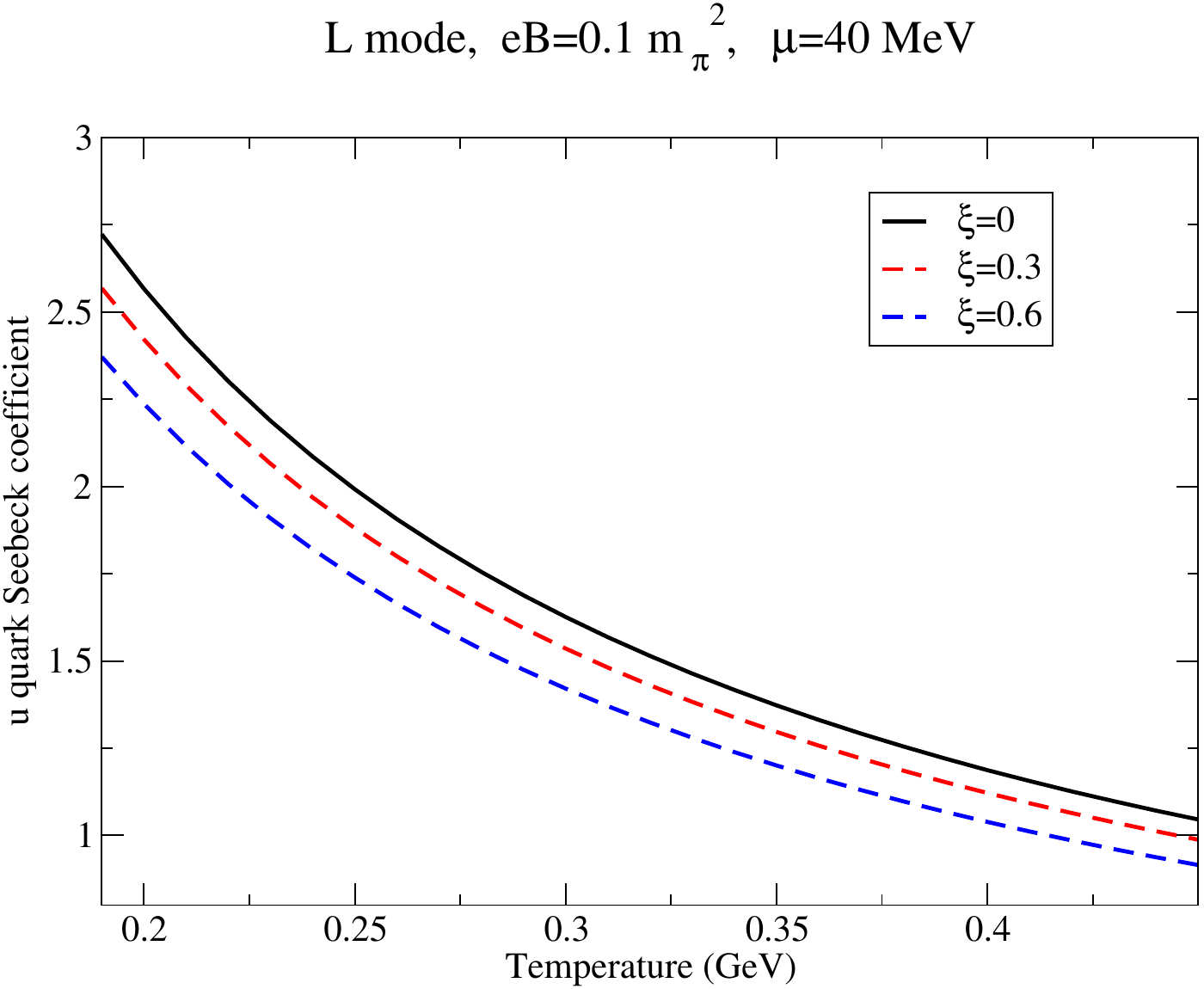}
		\caption{}\label{seebeck_B_u_L}
	\end{subfigure}
	\hspace*{\fill}
	\begin{subfigure}{0.48\textwidth}
		\includegraphics[width=0.95\textwidth,height=0.3\textheight]{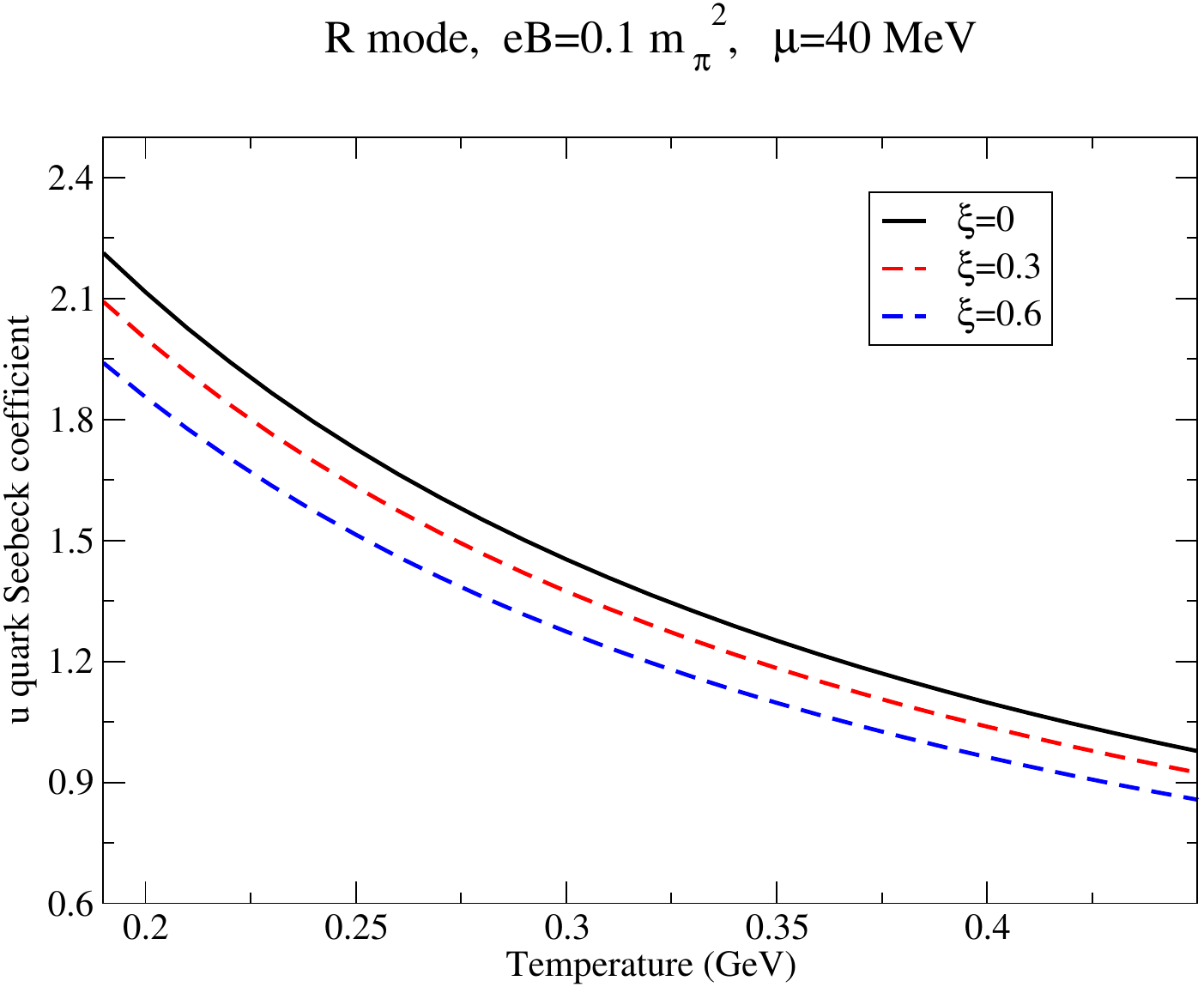}
		\caption{}\label{seebeck_B_u_R}
	\end{subfigure}
	\caption{(a) Temperature dependence of $u$ quark $L$ mode Seebeck coefficient at fixed values of $eB$ and $\mu$. (b) Temperature dependence of $u$ quark $R$ mode Seebeck coefficient at fixed values of $eB$ and $\mu$. The different curves correspond to different values of $\xi$ }
	\label{seebeck_B_u} 
\end{figure}

\begin{figure}
	\begin{subfigure}{0.48\textwidth}
		\includegraphics[width=0.95\textwidth,height=0.3\textheight]{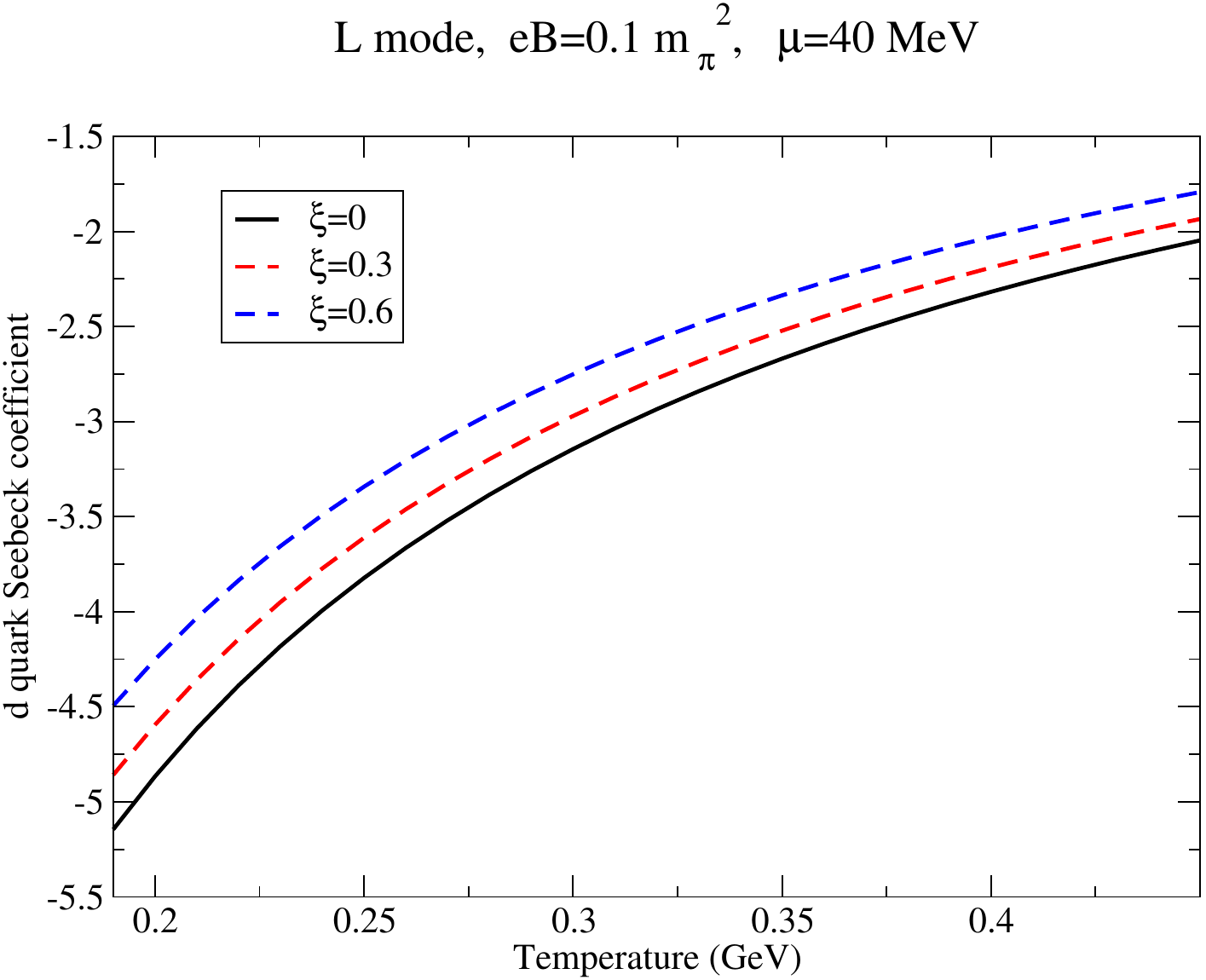}
		\caption{}\label{seebeck_B_d_L}
	\end{subfigure}
	\hspace*{\fill}
	\begin{subfigure}{0.48\textwidth}
		\includegraphics[width=0.95\textwidth,height=0.3\textheight]{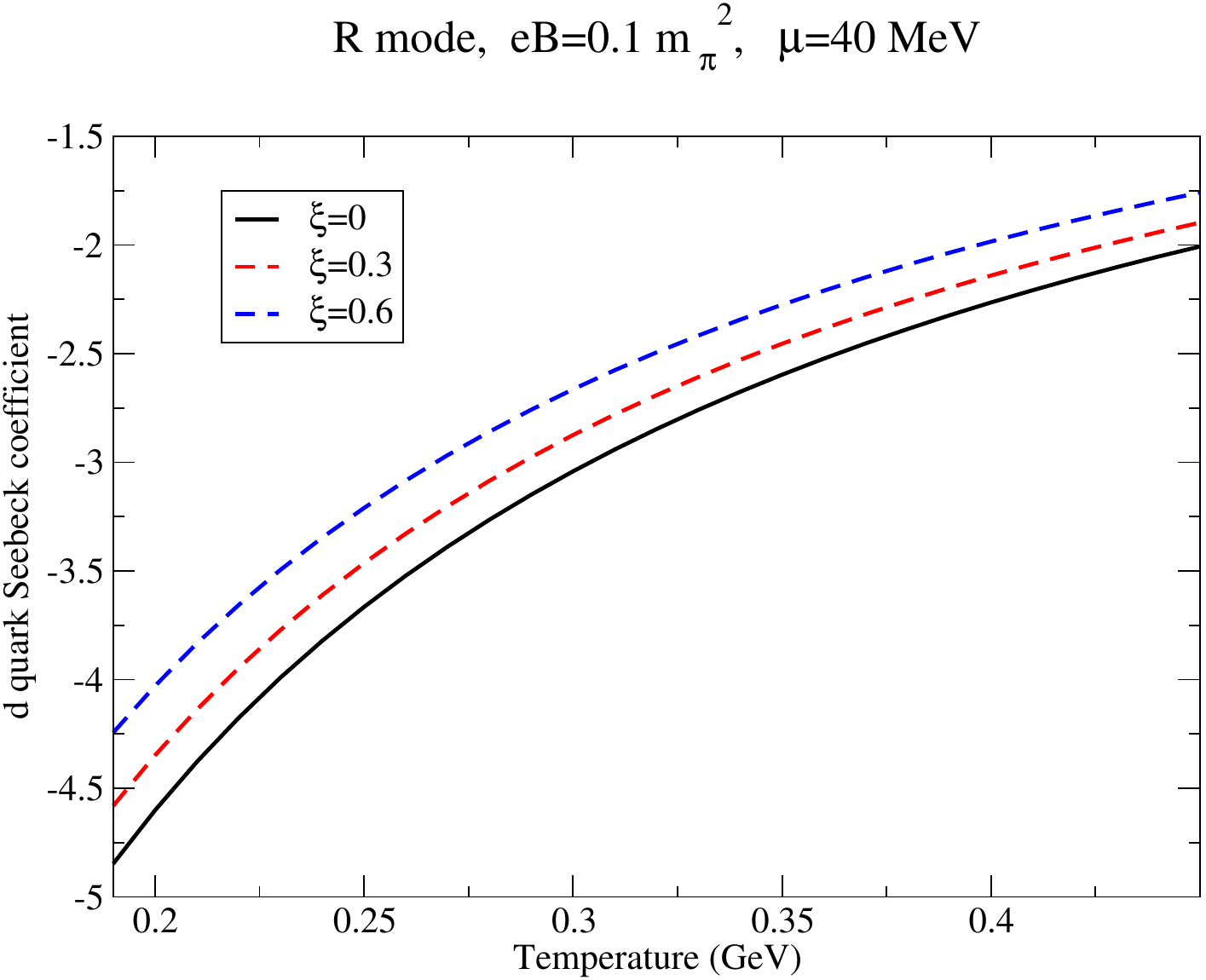}
		\caption{}\label{seebeck_B_d_R}
	\end{subfigure}
	\caption{(a) Temperature dependence of $d$ quark $L$ mode Seebeck coefficient at fixed values of $eB$ and $\mu$. (b) Temperature dependence of $d$ quark $R$ mode Seebeck coefficient at fixed values of $eB$ and $\mu$. The different curves correspond to different values of $\xi$.   }
	\label{seebeck_B_d} 
\end{figure}
Figures \eqref{seebeck_B_u} and \eqref{seebeck_B_d} show the variation with temperature of $u$ quark and $d$ quark Seebeck coefficients, respectively. The coefficient is positive for positively charged $u$ quark and negative for the negatively charged $d$ quark, which is along expected lines.  As can be seen from the figures, the coefficient magnitudes are decreasing functions of temperature. One can also see the effect of anisotropy on the coefficient magnitudes. Compared to the isotropic ($\xi=0$) result, the anisotropic medium leads to a  lesser value of the Seebeck coefficient magnitude for a particular value of temperature and magnetic field. Also, with the increase in the extent of anisotropy, the coefficient magnitude decreases. Comparing the graphs for the $L$ and $R$ modes shows that the Seebeck coefficient magnitudes for the $R$ mode are slightly smaller than its $L$ mode counterpart. This is due to the smaller effective mass of the $R$ mode compared to the $L$ mode at the same values of temperature and magnetic field. Specifically, the difference between the magnitudes of the $L$ and $R$ modes is greatest at lower temperatures and decreases as the temperature rises. We can take the average in the entire temperature range and define an average percentage change corresponding to each value of $\xi$. 
We found that the percentage decrease in the $u$ quark coefficient magnitude as one goes from the $L$ mode to the $R$ mode is $\approx$9.95\% for $\xi=0$, $\approx$9.86\% for $\xi=0.3$ and $\approx$9.68\% for $\xi=0.6$. This shows that as the strength of anisotropy increases, the difference between the $L$ and $R$ magnitudes decreases. Taking the mean of the three values corresponding to the three $\xi$ values, we arrive at a mean percentage decrease value of $\approx 9.83\%$. Interestingly, for the $d$ quark Seebeck coefficient, these values are $\approx3.06\%$ for $\xi=0$, $\approx 3.03\%$ for $\xi=0.3$ and $\approx 2.97\%$ for $\xi=0.6$. So, the effect of the difference in $L$ and $R$ mode quasiparticle masses is suppressed for the $d$ quark Seebeck coefficient compared to the $u$ quark result.

\vspace{4mm}
\begin{figure}
	\begin{subfigure}{0.48\textwidth}
		\includegraphics[width=0.95\textwidth,height=0.3\textheight]{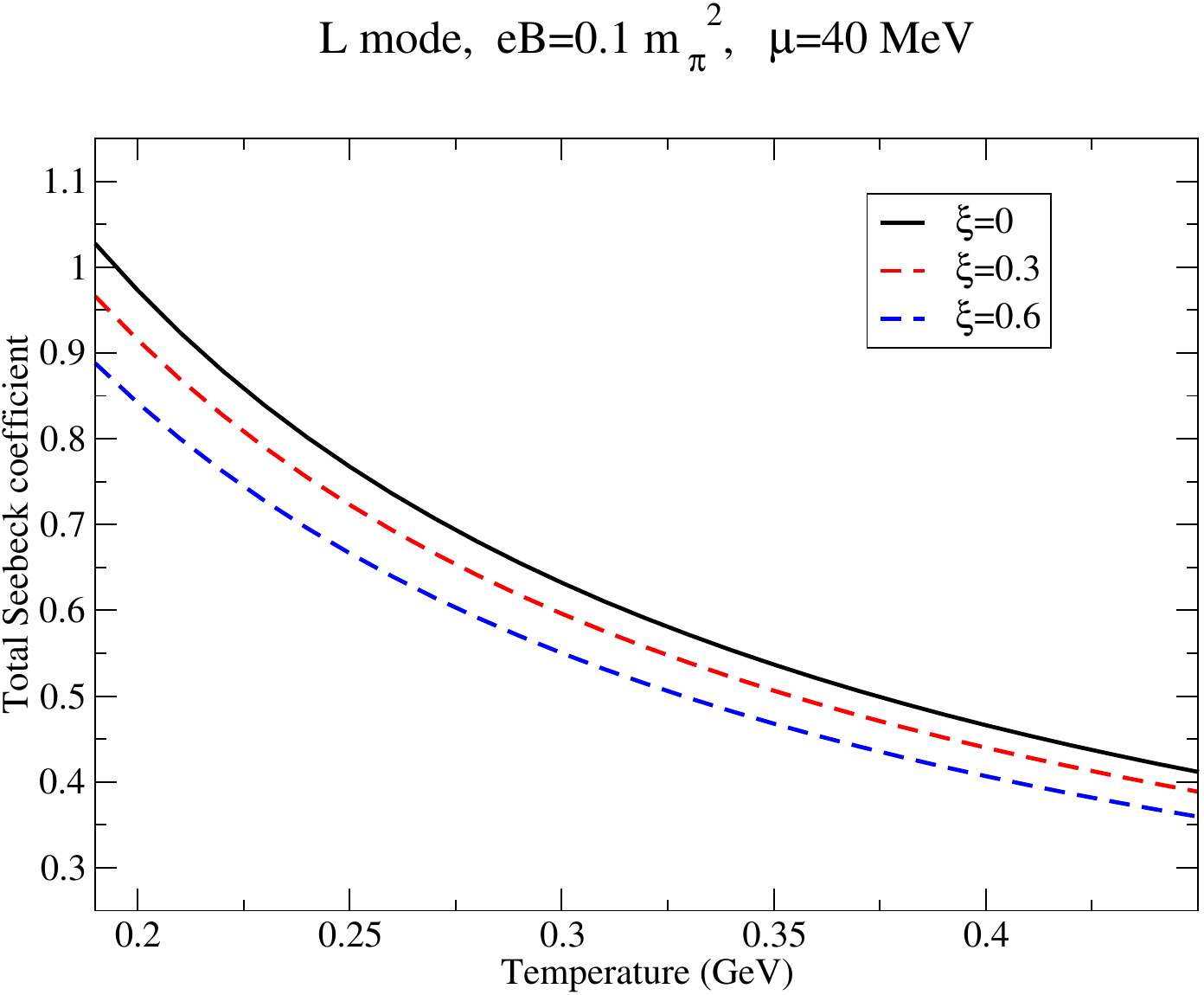}
		\caption{}\label{seebeck_B_med_L}
	\end{subfigure}
	\hspace*{\fill}
	\begin{subfigure}{0.48\textwidth}
		\includegraphics[width=0.95\textwidth,height=0.3\textheight]{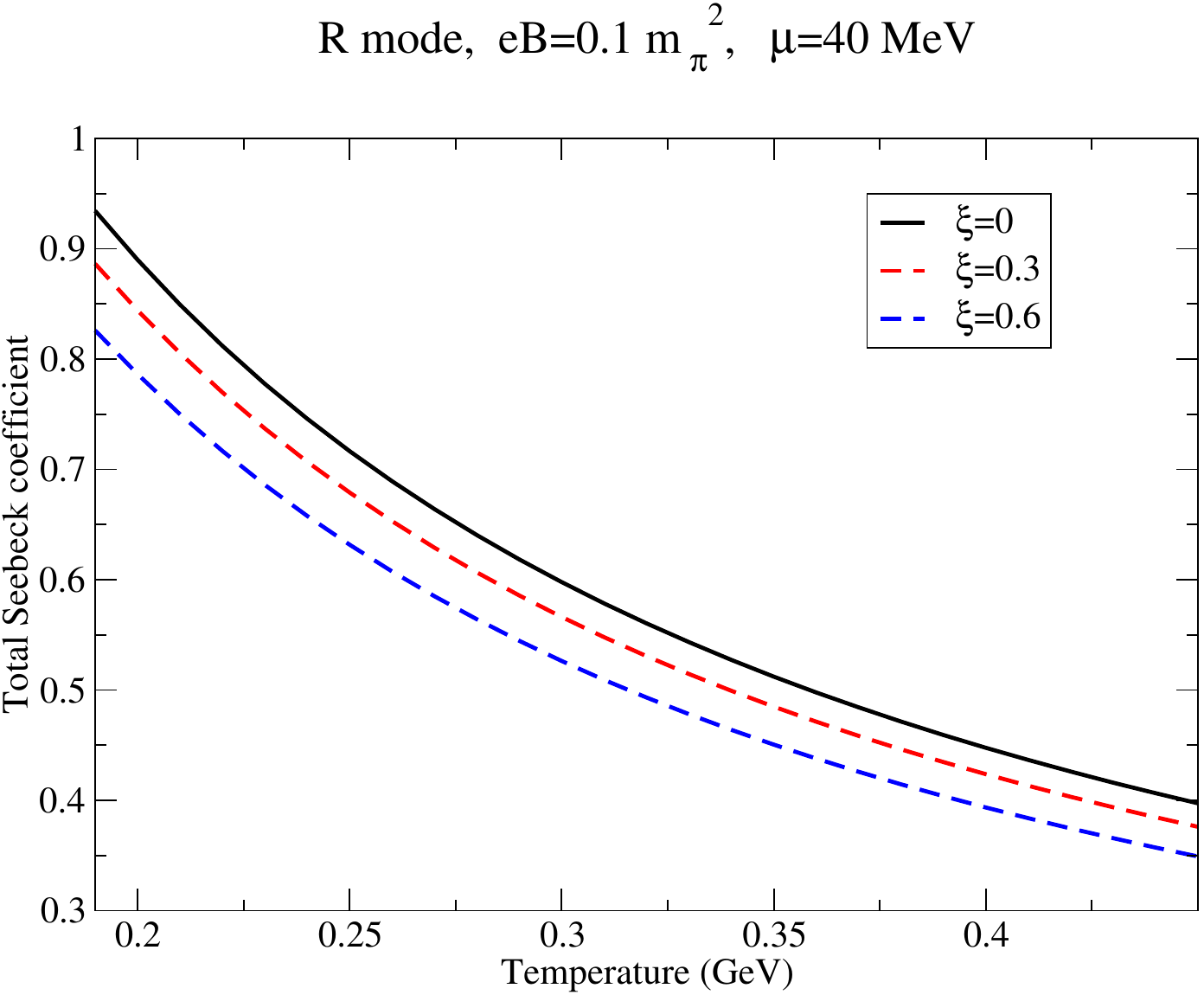}
		\caption{}\label{seebeck_B_med_R}
	\end{subfigure}
	\caption{(a) Variation of total Seebeck coefficient of the medium ($L$ mode) with temperature at fixed values of $eB$ and $\mu$. (b) Variation of total Seebeck coefficient of the medium ($R$ mode) with temperature at fixed values of $eB$ and $\mu$. The different curves correspond to different values of $\xi$.    }
	\label{seebeck_B_med} 
\end{figure}
Figures \eqref{seebeck_B_med_L} and \eqref{seebeck_B_med_R} show the temperature dependence of  Seebeck coefficient of the composite medium composed of $u$ and $d$ quarks. The total Seebeck coefficient is positive, which means that the induced electric field of the medium points in the direction of increasing temperature.  The coefficient magnitude decreases with temperature and also decreases with increase in the anisotropy parameter $\xi$. As in the case of the individual coefficients, the $R$ mode Seebeck coefficient magnitude is slightly smaller than that of the $L$ mode, which again can be attributed to the smaller effective mass of the $R$ mode quasiquark. The average (over $T$ and $\xi$ both) percentage decrease as one goes from the $L$ mode to the $R$ mode composite Seebeck coefficient is $\approx 4.61\%$. Here also, the percentage difference between the two modes decreases with increasing anisotropy strength. Thus, anisotropic expansion of the medium hinders the ability of a thermal QCD medium to convert a temperature gradient to electric field.

\vspace{4mm}
\begin{figure}
	\begin{subfigure}{0.48\textwidth}
		\includegraphics[width=0.95\textwidth,height=0.3\textheight]{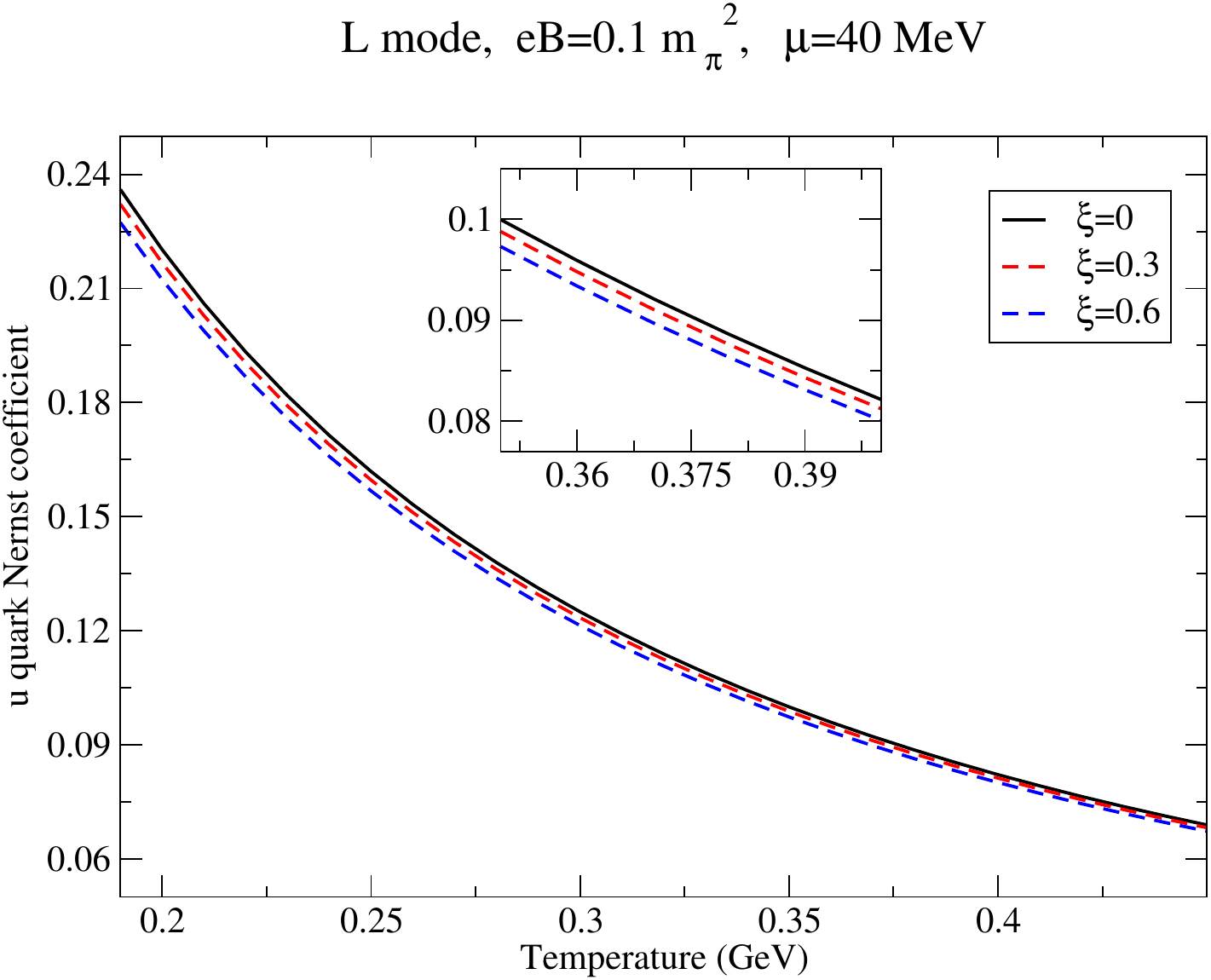}
		\caption{}\label{nernst_u_L}
	\end{subfigure}
	\hspace*{\fill}
	\begin{subfigure}{0.48\textwidth}
		\includegraphics[width=0.95\textwidth,height=0.3\textheight]{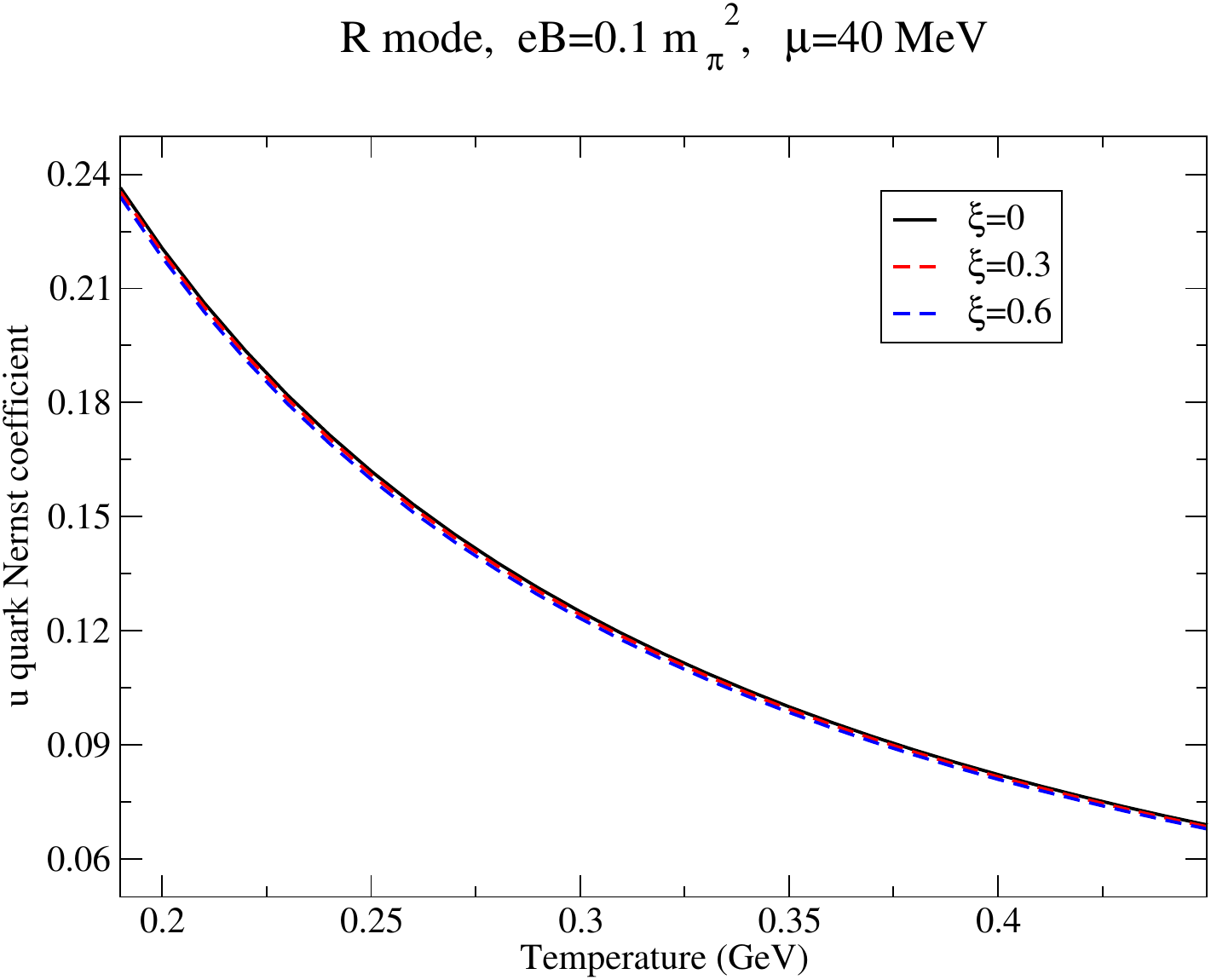}
		\caption{}\label{nernst_u_R}
	\end{subfigure}
	\caption{(a) Variation of $u$ quark $L$ mode Nernst coefficient with temperature at fixed values of $eB$ and $\mu$. (b) Variation of $u$ quark $R$ mode Nernst coefficient with temperature at fixed values of $eB$ and $\mu$. The different curves correspond to different values of $\xi$.}
	\label{nernst_u} 
\end{figure}
\begin{figure}
	\begin{subfigure}{0.48\textwidth}
		\includegraphics[width=0.95\textwidth,height=0.3\textheight]{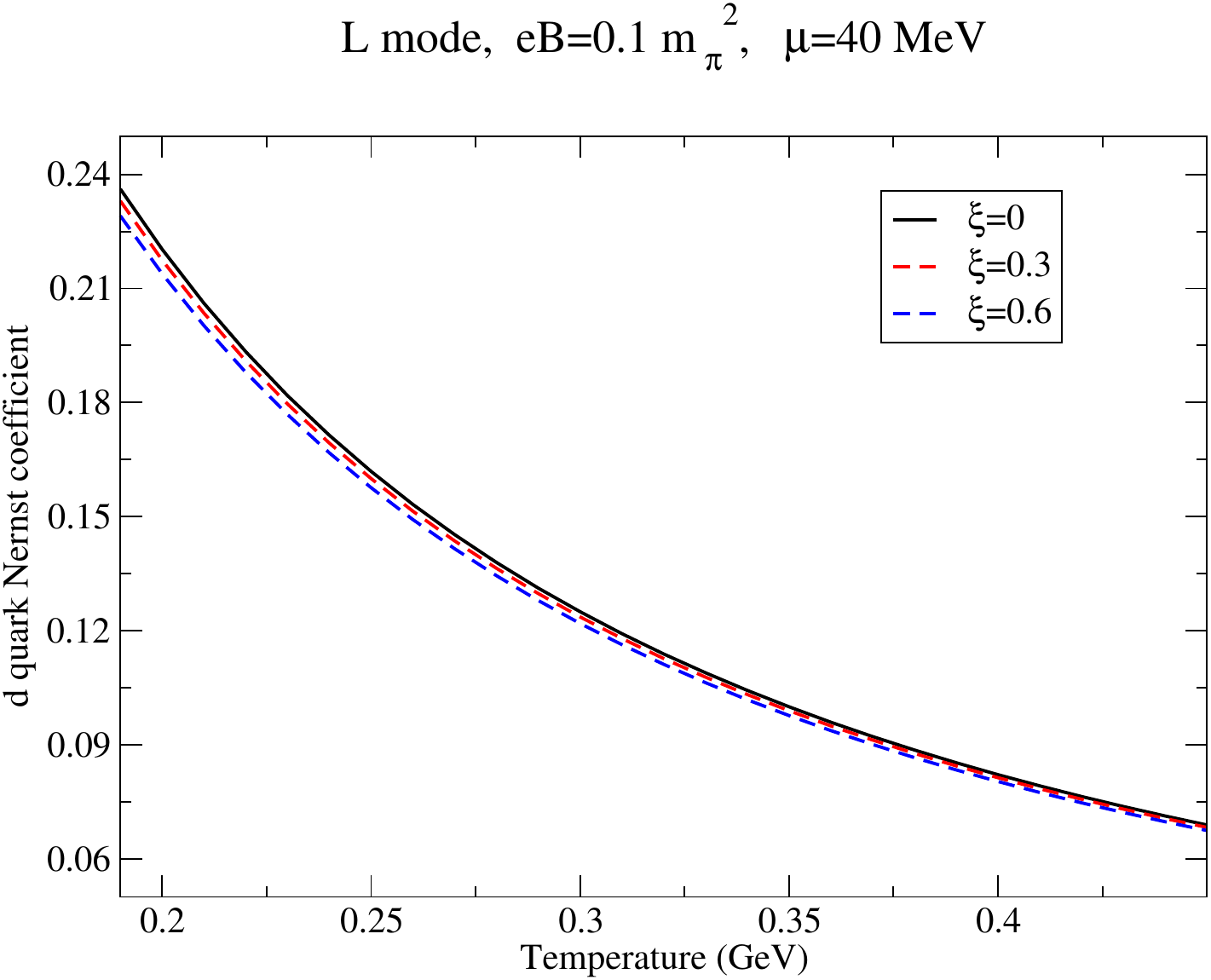}
		\caption{}\label{nernst_d_L}
	\end{subfigure}
	\hspace*{\fill}
	\begin{subfigure}{0.48\textwidth}
		\includegraphics[width=0.95\textwidth,height=0.3\textheight]{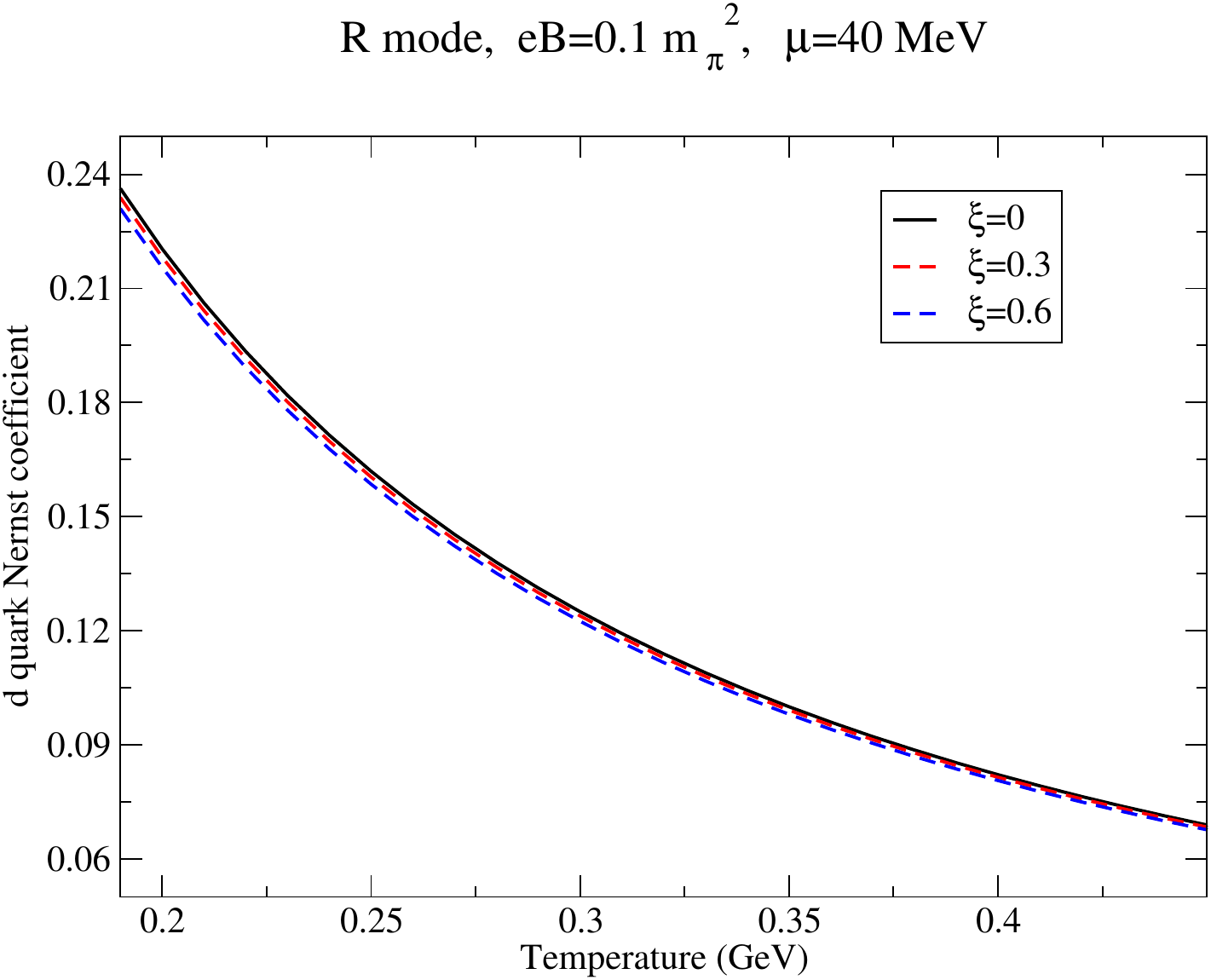}
		\caption{}\label{nernst_d_R}
	\end{subfigure}
	\caption{(a) Variation of $d$ quark $L$ mode Nernst coefficient with temperature at fixed values of $eB$ and $\mu$. (b) Variation of $d$ quark $R$ mode Nernst coefficient with temperature at fixed values of $eB$ and $\mu$. The different curves correspond to different values of $\xi$..}
	\label{nernst_d} 
\end{figure}
Figures \eqref{nernst_u_L} and \eqref{nernst_u_R} show the variation with temperature, of the Nernst coefficient corresponding to the $L$ and $R$ modes, respectively, of a medium composed exclusively of $u$ quarks. Similar to the Seebeck coefficient, the Nernst coefficient magnitude decreases with temperature and also decreases with the value of the anisotropy parameter $\xi$. Comparison between the graphs corresponding to the $L$ and $R$ modes reveals that $R$ mode magnitudes are slightly more than that of the $L$ mode. This trend is opposite to the individual Seebeck coefficient case where $R$ mode magnitudes were less than their $L$ mode counterparts.  Also, the extent of difference between the two modes is much smaller compared to the Seebeck coefficient case. Specifically, for the $u$ quark Nernst coefficient, the temperature averaged percentage decrease as one goes from $R$ mode to $L$ mode is $0.051\%$ for $\xi=0$, $\approx 0.65\%$ for $\xi=0.3$ and $\approx 1.43\%$ for $\xi=0.6$. Taking the mean of the values corresponding to the different $\xi$ values gives us an average value of $0.71\%$. Compared to the $u$ quark Seebeck coefficient, these values are almost an order of magnitude smaller.
Also, unlike in the case of the Seebeck coefficients, the percentage change between the $L$ and $R$ mode increases sharply with increase in the strength of anisotropy,

The $d$ quark Nernst coefficients corresponding to the $L$ and $R$ modes shown in Figs.\eqref{nernst_d_L} and \eqref{nernst_d_R} respectively show that the magnitudes of Nernst coefficients corresponding to both the modes decreases with temperature. Also, the magnitudes decrease with increase in the degree of anisotropy, parameterized by the value of $\xi$. 
Similar to the $u$ quark Nernst coefficient, the $L$ mode absolute values of $d$ quark Nernst coefficients are smaller than their $R$ mode counterparts; the averaged (over $T$) percentage decrease being $0.015\%$ for $\xi=0$, $\approx 0.192\%$ for $\xi=0.3$ and $\approx 0.422\%$ for $\xi=0.6$. Averaging also over the different $\xi$ values yields a mean percentage decrease value of $0.21\%$. This trend is similar to the individual Seebeck coefficient case where the percentage changes for $d$ quark Seebeck coefficient was much smaller than that for the $u$ quark.   Also from the magnitudes of the individual Seebeck and Nernst coefficients (both $u$ and $d$ quarks), it can be seen that magnitudes of the Nernst coefficients are $\approx$ 1 order of magnitude smaller.

 The major point of difference with the Seebeck coefficient, however, is that the Nernst coefficient for both $u$ and $d$ quarks is positive. To understand this, let us consider positively charged quarks moving in the $+\hat{x}$ direction under the influence of a temperature gradient. On application of a magnetic field in the $\hat{z}$ direction, the Lorentz force will cause them to drift in the $-\hat{y}$ direction. This will result in an induced electric field in the $+\hat{y}$ direction. If the electric charge of the quarks were negative instead, they would drift towards the $+\hat{y}$ direction and pile up there. This would again lead to an induced electric field in the $+\hat{y}$ direction. Thus, the direction of the induced field does not depend on the sign of the electric charge of the quark, unlike in the case of Seebeck coefficient, resulting in positive Nernst coefficients for both $u$ and $d$ quarks.

\begin{figure}
	\begin{subfigure}{0.48\textwidth}
		\includegraphics[width=0.95\textwidth,height=0.3\textheight]{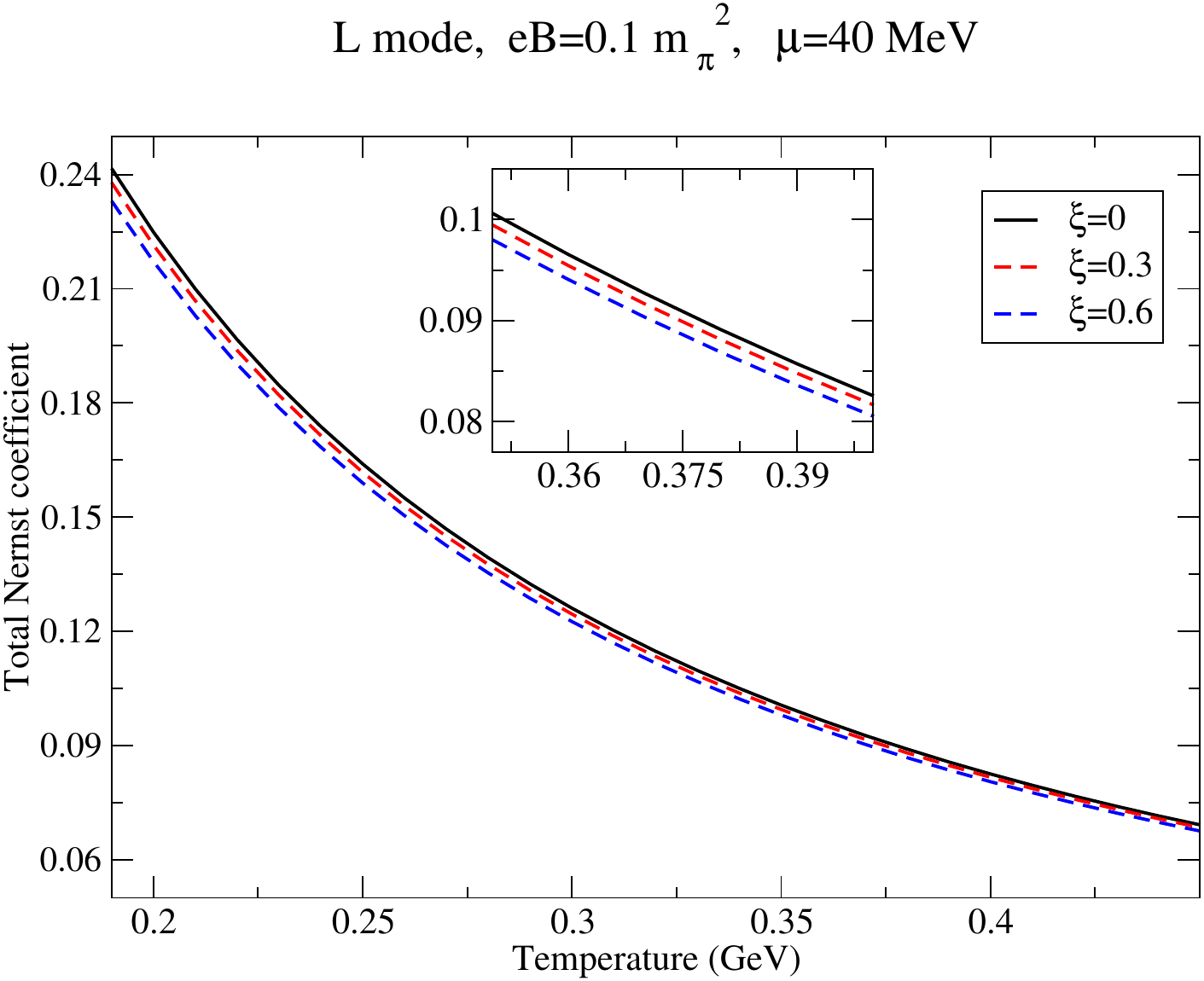}
		\caption{}\label{nernst_med_L}
	\end{subfigure}
	\hspace*{\fill}
	\begin{subfigure}{0.48\textwidth}
		\includegraphics[width=0.95\textwidth,height=0.3\textheight]{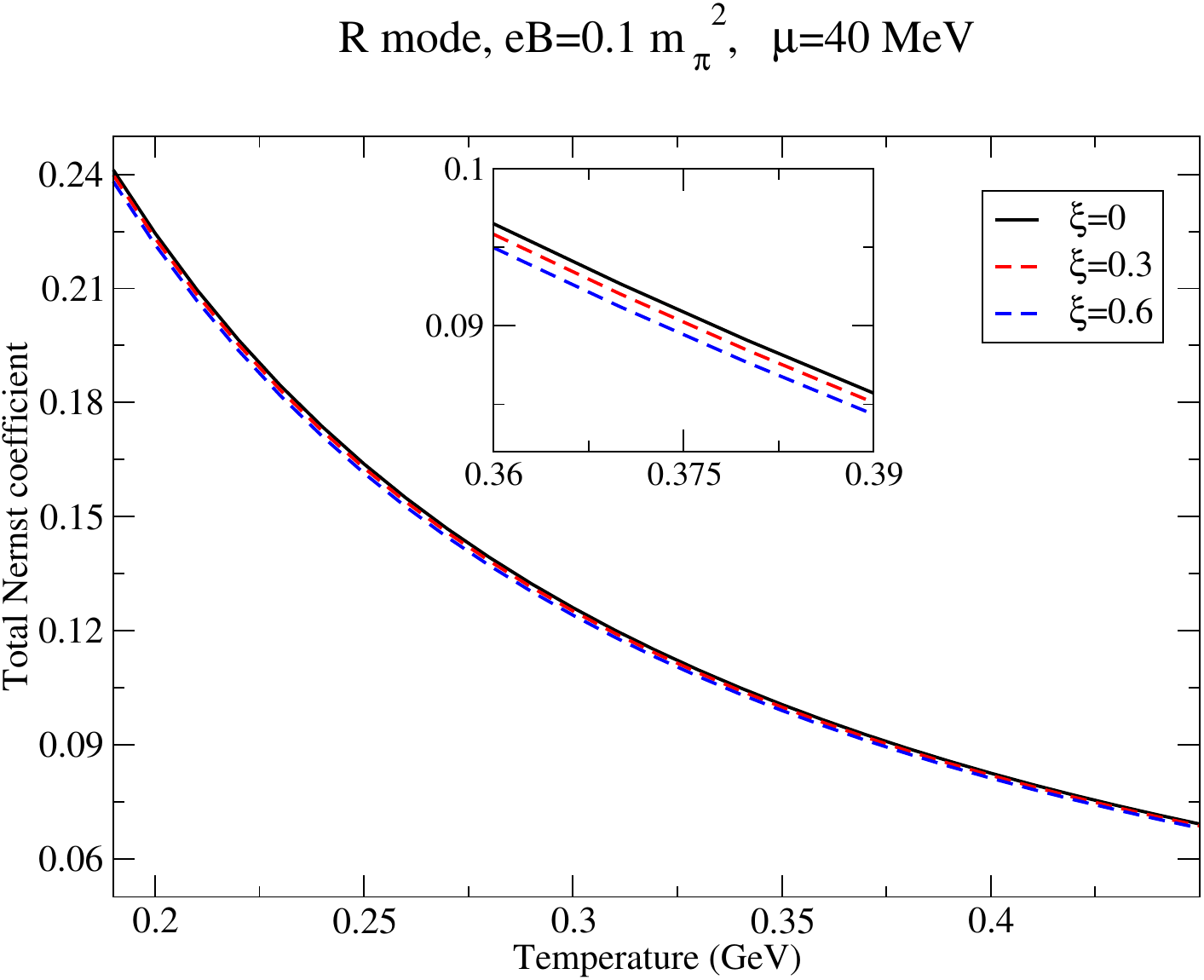}
		\caption{}\label{nernst_med_R}
	\end{subfigure}
	\caption{(a) Variation of total Nernst coefficient of the medium ($L$ mode) with temperature at fixed values of $eB$ and $\mu$. (b) Variation of total Nernst coefficient of the medium ($R$ mode) with temperature at fixed values of $eB$ and $\mu$. The different curves correspond to different values of $\xi$.}
	\label{nernst_med} 
\end{figure}
Figures \eqref{nernst_med_L} and \eqref{nernst_med_R} show the temperature variation of the Nernst coefficients of the composite medium corresponding to $L$ mode quasiparticles and $R$ mode quasiparticles, respectively. The total Nernst coefficient of the medium is positive, and is a decreasing function of temperature. Like its Seebeck counterpart, its magnitude  decreases with the strength of anisotropy. Compared to the individual Nernst coefficient magnitudes, the total Nernst coefficient magnitudes are an order of magnitude smaller. Also, compared to the total Seebeck coefficient, its values are an order of magnitude smaller. This suggests that the Nernst effect in a weakly magnetized thermal QCD medium is a weaker effect compared to the Seebeck effect.  Just as in the case of the individual Nernst coefficients, the $L$ mode values of the total Nernst coefficient are  slightly smaller in magnitude than their $R$ mode counterparts; the averaged (over $T$) percentage decrease being $0.04\%$ for $\xi=0$, $\approx 0.47\%$ for $\xi=0.3$ and $\approx 1.14\%$ for $\xi=0.6$. The mean of the three values corresponding to different $\xi$ is $0.55\%$. Interestingly, these numbers are $\approx$ an order of magnitude bigger than corresponding numbers for the individual Nernst coefficients, and an order of magnitude smaller than the corresponding values for the total Seebeck coefficient. Such a drastic change in going from the individual to the total coefficients was not observed for the Seebeck coefficient. 
Thus, the sensitivity to the mass difference between the $L$ and $R$ modes of quarks is much amplified for the composite medium, compared to a single flavor medium.

\section{Conclusion}
We have estimated the thermoelectric response of a deconfined hot QCD medium in the presence of a weak external magnetic field, taking into account the anisotropic expansion of the QGP fireball. The strength of the response, \textit{i.e.} the ability to convert a temperature gradient into an electric field, is quantified by two coefficients, \textit{viz.} Seebeck coefficient and Nernst coefficient. We have calculated the individual as well as the total response coefficients of the medium and checked their variation with temperature and anisotropy strength. We have presented results both in the absence and presence of a background magnetic field $B$. Our plots for finite $B$ have been generated for a constant background magnetic field of strength $eB=0.1\,m_{\pi}^2$, and a constant chemical potential $\mu=40 MeV$. 
We have found that the magnitudes of both the individual as well as the total coefficients-both Seebeck and Nernst, are decreasing functions of temperature, and decreasing functions of anisotropy strength, characterized by the anisotropy parameter $\xi$. It is important to note that the Seebeck coefficient vanishes for $\mu=0$ irrespective of the magnetic field strength, and the Nernst coefficient vanishes for $|\bm{B}|=0$, irrespective of the value of $\mu$. We have analysed the sensitivity to the $L$ and $R$ modes, \textit{i.e.} to the difference in quasiparticle effective masses of the coefficient magnitudes; both for individual and total coefficients. To that end we have calculated the average percentage change in the coefficient magnitude as one goes from the $L$ mode to the $R$ mode or vice versa. For the Seebeck coefficient, the average (over $T$ and $\xi$) percentage change for the $u$ quark is $\approx9.83\%$, whereas for the $d$ quark, it is $\approx 3.02\%$. This shows that the $d$ quark Seebeck coefficient is comparatively less sensitive to differences in quasiparticle masses. For the total Seebeck coefficient, the average percentage change in going from the $L$ mode to the $R$ mode is $\approx4.61\%$, which is in between the values for the individual coefficients.

Certain differences arise in the case of Nernst coefficient. Firstly, for both the individual and total Nernst coefficients, the absolute values of the $R$ mode coefficients are greater than that of the $L$ mode. This is opposite to the case of Seebeck coefficients. The average percentage change in going from the $R$ mode to the $L$ mode in case of $u$ quark Nernst coefficient is $\approx 0.71\%$; for the $d$ quark Nernst coefficient, this value is $\approx0.21\%$. Compared to their Seebeck counterparts, these numbers are an order of magnitudes smaller, indicating that the individual Nernst coefficients are comparatively much less sensitive to the change in quasiparticle modes. A major difference between the individual Nernst and Seebeck coefficients is that while positively (negatively) charged quarks lead to positive (negative) Seebeck coefficients, the individual Nernst coefficients are independent of the electric charge of the quark. For the case of the total Nernst coefficient, the average percentage decrease from $R$ to $L$ modes is $\approx0.55\%$. 
A nonzero value of the Seebeck coefficient will modify the
electric, as well as heat, current in the medium. So a better understanding of the thermoelectric transport phenomena in the 
medium is necessary to estimate the life time of the magnetic field 
and phenomenology of the QGP. 

\section{Acknowledgements}
SAK is thankful to the authorities of Integral University for providing the necessary facilities for research and assigning the manuscript communication number- IU/R\&D/2024-MVN-0002668.



\begin{thebibliography}{9}
\bibitem{Shuryak:PLA757'1978}E. V. Shuryak, \href{https://www.sciencedirect.com/science/article/abs/pii/0370269378903702?via%3Dihub}{  Phys. Lett. A \textbf{78}, 1 (1978).}
	
\bibitem{Kapusta:PRD44'1991}J. I. Kapusta, P. Lichard, and D. Seibert, \href{https://journals.aps.org/prd/abstract/10.1103/PhysRevD.44.2774}{Phys. Rev. D \textbf{44}, 2774 (1991).}

\bibitem{Blaizot:PRL77'1996}J. P. Blaizot and J. Y. Ollitrault, \href{https://journals.aps.org/prl/abstract/10.1103/PhysRevLett.77.1703}{Phys. Rev. Lett. \textbf{77}, 1703 (1996).}

\bibitem{Satz:NPA783'2007}H. Satz, \href{https://www.sciencedirect.com/science/article/pii/S0375947406008256?via%3Dihub}{Nucl. Phys. A \textbf{783}, 249 (2007).}
	
\bibitem{Rapp:PPNP65'2010}R. Rapp, D. Blaschke, and P. Crochet, \href{https://www.sciencedirect.com/science/article/pii/S0146641010000505?via%3Dihub}{Prog. Part. Nucl. Phys. \textbf{65}, 209 (2010).}
	
\bibitem{Bhalerao:PLB641'2006}R. S. Bhalerao and J. Y. Ollitrault, \href{https://www.sciencedirect.com/science/article/pii/S0370269306010318?via%3Dihub}{Phys. Lett. B \textbf{641}, 260 (2006).}
	
\bibitem{Voloshin:PLB659'2008} S. A. Voloshin, A. M. Poskanzer, A. Tang, and G. Wang,
\href{https://www.sciencedirect.com/science/article/pii/S0370269307014542?via%3Dihub}{Phys. Lett. B 659, 537 (2008).}
	
\bibitem{Wang:PRL68'1992}X. N. Wang and M. Gyulassy, \href{https://journals.aps.org/prl/abstract/10.1103/PhysRevLett.68.1480}{Phys. Rev. Lett. \textbf{68}, 1480 (1992).}

\bibitem{Adcox:PRL88'2001}K. Adcox et al., \href{https://journals.aps.org/prl/abstract/10.1103/PhysRevLett.88.022301}{Phys. Rev. Lett. \textbf{88}, 022301 (2001).}	

\bibitem{Chatrchyan:PRC84'2011}S. Chatrchyan et al., \href{https://journals.aps.org/prc/abstract/10.1103/PhysRevC.84.024906}{Phys. Rev. C \textbf{84}, 024906 (2011).}

\bibitem{Arsene:NuclPhysA757'2005}I. Arsene et al., BRAHMS Collaboration, \href{https://www.sciencedirect.com/science/article/pii/S0375947405002770}{ Nucl. Phys. A \textbf{757}, 1-27 (2005).}

\bibitem{Adams:NuclPhysA757'2005}J. Adams et al., STAR Collaboration, \href{https://www.sciencedirect.com/science/article/pii/S0375947405005294}{Nucl. Phys. A \textbf{757},  102-183 (2005)}.

\bibitem{Adcox:NuclPhysA757'2005}K. Adcox et al., PHENIX Collaboration, \href{https://www.sciencedirect.com/science/article/pii/S0375947405005300}{Nucl. Phys. A \textbf{757}, 184-283 (2005).}

\bibitem{Carminati:JPhysG30'2004}F. Carminati et al., ALICE Collaboration, \href{https://iopscience.iop.org/article/10.1088/0954-3899/30/11/001}{J. Phys. G: Nucl. Part. Phys. \textbf{30}, 1517 (2004).}

\bibitem{Alessandro:JPhysG30'2006}B. Alessandro et al., ALICE Collaboration, \href{https://iopscience.iop.org/article/10.1088/0954-3899/32/10/001}{J. Phys. G: Nucl. Part. Phys. \textbf{32}, 1295 (2006)}.

\bibitem{Aoki:PLB643'2006}Y. Aoki, Z. Fodor, S. D. Katz and K. K. Szabo, \href{https://www.sciencedirect.com/science/article/pii/S0370269306012755?via%3Dihub}{ Phys. Lett. B \textbf{643}, 46 (2006).}	
	
\bibitem{Borsanyi:JHEP1009'2010}S. Borsanyi \textit{et al.} \href{https://link.springer.com/article/10.1007/JHEP09(2010)073}{JHEP \textbf{1009}, 073 (2010).}
	
\bibitem{Borsanyi:PRD92'2015}S. Borsanyi \textit{et al.}, \href{https://journals.aps.org/prd/abstract/10.1103/PhysRevD.92.014505}{Phys. Rev. D \textbf{92}, 014505 (2015).}

\bibitem{Ding:IntJModPhysE24'2015}H. T. Ding, F. Karsch and S. Mukherjee, \href{https://www.worldscientific.com/doi/abs/10.1142/S0218301315300076}{Int. J. Mod. Phys E \textbf{24}, 1530007 (2015).}

\bibitem{Hands:NuclPhysB106-107'2002}S. Hands, \href{https://www.sciencedirect.com/science/article/pii/S092056320101653X}{Nucl. Phys. B \textbf{106-107} (2002).}

\bibitem{Alford:NuclPhysProcSuppl117'2003}M. G. Alford, \href{https://www.sciencedirect.com/science/article/pii/S0920563203014117}{Nucl. Phys. Proc. Suppl. \textbf{117}, 65 (2003).}

\bibitem{Romatschke:PRD68'2003}P. Romatschke and M. Strickland, \href{https://journals.aps.org/prd/abstract/10.1103/PhysRevD.68.036004}{Phys. Rev. D \textbf{68}, 036004 (2003).}

\bibitem{Romatschke:PRD70'2004}P. Romatschke and M. Strickland, \href{https://journals.aps.org/prd/abstract/10.1103/PhysRevD.68.036004}{Phys. Rev. D \textbf{70}, 116006 (2004).}

\bibitem{Carrington:PRC90'2014}M. E. Carrington, K. Deja, and S. Mrówczyński, \href{https://journals.aps.org/prc/abstract/10.1103/PhysRevC.90.034913}{Phys. Rev. C \textbf{90}, 034913 (2014).}

\bibitem{Schenke:PRD74'2006}B. Schenke and M. Strickland, \href{https://journals.aps.org/prd/abstract/10.1103/PhysRevD.74.065004}{Phys. Rev. D \textbf{74}, 065004 (2006).}

\bibitem{Thoma:PRD62'2000}S. Mr\'{o}wczy\'{n}ski and M. H. Thoma, \href{}{Phys. Rev. D \textbf{62}, 036011 (2000).}

\bibitem{Randrup:PRC68'2003}J. Randrup and S. Mr\'{o}wczy\'{n}ski, \href{https://journals.aps.org/prc/abstract/10.1103/PhysRevC.68.034909}{Phys. Rev. C \textbf{68}, 034909 (2003)}.

\bibitem{Strickland:PREP682'2017}S. Mr\'{o}wczy\'{n}ski, B. Schenke, and M. Strickland, \href{https://linkinghub.elsevier.com/retrieve/pii/S0370157317300571}{Phys. Rep. \textbf{682}, 1 (2017).}

\bibitem{Schenke:PRD76'2007}B. Schenke and M. Strickland, \href{https://journals.aps.org/prd/abstract/10.1103/PhysRevD.76.025023}{Phys. Rev. D \textbf{76}, 025023 (2007).}

\bibitem{Bhattacharya:PRD93'2016}L. Bhattacharya, R. Ryblewski, and M. Strickland, \href{https://journals.aps.org/prd/abstract/10.1103/PhysRevD.93.065005}{Phys. Rev. D \textbf{93}, 065005 (2016).}

\bibitem{Nopoush:JHEP09'2017}M. Nopoush, Y. Guo, and M. Strickland, \href{https://link.springer.com/article/10.1007/JHEP09(2017)063}{J. High Energy
	Phys. \textbf{09} (2017) 063.}

\bibitem{Krouppa:UNI2'2016}B. Krouppa and M. Strickland, \href{https://www.mdpi.com/2218-1997/2/3/16}{Universe \textbf{2}, 16 (2016).}

\bibitem{Kasmaei:PRD97'2018}B. S. Kasmaei and M. Strickland\href{https://journals.aps.org/prd/abstract/10.1103/PhysRevD.97.054022}{Phys. Rev. D \textbf{97}, 054022 (2018)}

\bibitem{Florkowski:PRC83'2011}W. Florkowski and R. Ryblewski, \href{https://journals.aps.org/prc/abstract/10.1103/PhysRevC.83.034907}{Phys. Rev. C \textbf{83}, 034907 (2011).}

\bibitem{Martinez:NPA848'2010} M. Martinez and M. Strickland, \href{https://www.sciencedirect.com/science/article/pii/S0375947410006536?via%3Dihub}{Nucl. Phys. A \textbf{848}, 183 (2010).}
	
\bibitem{Alqahtani:PRC96'2017}M. Alqahtani, M. Nopoush, R. Ryblewski, and M.
Strickland, \href{https://journals.aps.org/prc/abstract/10.1103/PhysRevC.96.044910}{Phys. Rev. C \textbf{96}, 044910 (2017).}
	
\bibitem{Alqahtani:PPNP101'2018}M. Alqahtani, M. Nopoush, and M. Strickland, \href{https://www.sciencedirect.com/science/article/pii/S0146641018300371?via%3Dihub}{Progress in Particle and Nuclear Physics \textbf{101}, 204-208 (2018)}.
		
\bibitem{Alqahtani:PRL119'2017}M. Alqahtani, M. Nopoush, R. Ryblewski, and M. Strickland,\href{https://journals.aps.org/prl/abstract/10.1103/PhysRevLett.119.042301}{Phys. Rev. Lett. \textbf{119}, 042301 (2017).}

\bibitem{Florkowski:PRC89'2014}W. Florkowski, E. Maksymiuk, R. Ryblewski, and M.
Strickland, \href{https://journals.aps.org/prc/abstract/10.1103/PhysRevC.89.054908}{Phys. Rev. C \textbf{89}, 054908 (2014).}

\bibitem{Nopoush:PRC90'2014}M. Nopoush, R. Ryblewski, and M. Strickland, \href{https://journals.aps.org/prc/abstract/10.1103/PhysRevC.90.014908}{Phys. Rev. C \textbf{90}, 014908 (2014).}

\bibitem{Tinti:NPA946'2016}L. Tinti, R. Ryblewski, W. Florkowski, and M. Strickland,
\href{https://www.sciencedirect.com/science/article/pii/S0375947415002638?via%3Dihub}{Nucl. Phys. A \textbf{946}, 29 (2016).}
	
\bibitem{Strickland:NPA956'2016} M. Strickland, M. Nopoush, and R. Ryblewski, \href{https://www.sciencedirect.com/science/article/pii/S0375947416001019?via%3Dihub}{Nucl. Phys. A \textbf{956}, 268 (2016).}
		
\bibitem{Strickland:PRD97'2018}M. Strickland, J. Noronha, and G. Denicol, \href{https://journals.aps.org/prd/abstract/10.1103/PhysRevD.97.036020}{Phys. Rev. D \textbf{97}, 036020 (2018).}
		
\bibitem{Strickland:JHEP12'2018}M. Strickland, \href{https://link.springer.com/article/10.1007/JHEP12(2018)128}{J. High Energy Phys. \textbf{12} (2018) 128.}
		
\bibitem{Rath:PRD100'2019}S. Rath and B. K. Patra, \href{https://journals.aps.org/prd/abstract/10.1103/PhysRevD.100.016009}{Phys. Rev. D \textbf{100}, 016009 (2019)}
		
\bibitem{Kumar:PRC105'2022}A. Kumar, M. Kurian, S. K. Das and V. Chandra \href{https://journals.aps.org/prc/abstract/10.1103/PhysRevC.105.054903}{Phys. Rev. C \textbf{105}, 054903 (2022)}

\bibitem{Tuchin:AdvHEP'2013} K. Tuchin, \href{https://www.hindawi.com/journals/ahep/2013/490495/}{Adv.High Energy Phys. , 490495 (2013)}.

\bibitem{Sokov:IntJModPhysA24'2009} V. Skokov, A. Illarionov, and V. Toneev, \href{https://www.worldscientific.com/doi/abs/10.1142/S0217751X09047570}{Int. J. Mod. Phys. A \textbf{24}, 5925 (2009)}.

\bibitem{Tuchin:PRC82'2010}K. Tuchin, \href{https://journals.aps.org/prc/abstract/10.1103/PhysRevC.82.034904}{Phys. Rev. C \textbf{82}, 034904 (2010)}.

\bibitem{Tuchin:PRC83'2011}K. Tuchin, \href{https://journals.aps.org/prc/abstract/10.1103/PhysRevC.83.017901}{Phys. Rev. C \textbf{83}, 017901 (2011)}.

\bibitem{Marty:PRC88'2013}R. Marty, E. Bratkovskaya, W. Cassing, J. Aichelin and H. Berrehrah, \href{https://journals.aps.org/prc/abstract/10.1103/PhysRevC.88.045204}{Phys. Rev. C \textbf{88}, 045204 (2013)}.

\bibitem{Ding:PRD83'2011}H.-T. Ding, A. Francis, O. Kaczmarek, F. Karsch, E. Laermann, and W. Soeldner, \href{https://journals.aps.org/prd/abstract/10.1103/PhysRevD.83.034504}{Phys. Rev. D \textbf{83}, 034504  (2011)}.

\bibitem{Gupta:PLB597'2004}S. Gupta, \href{https://www.sciencedirect.com/science/article/pii/S0370269304008913?via%3Dihub}{Phys. Lett. B \textbf{597}, 57–62  (2004)}.
	
\bibitem{Amato:PRL111'2013}A. Amato, G. Aarts, C. Allton, P. Giudice, S. Hands, and J.-I. Skullerud, \href{https://journals.aps.org/prl/abstract/10.1103/PhysRevLett.111.172001}{Phys. Rev. Lett. \textbf{111}  no. 17, 172001 (2013)}. 

\bibitem{Aarts:PRL99'2007}G. Aarts, C. Allton, J. Foley, S. Hands, and S. Kim, \href{https://journals.aps.org/prl/abstract/10.1103/PhysRevLett.99.022002}{Phys. Rev. Lett. \textbf{99}, 022002  (2007)}. 

\bibitem{Puglisi:PRD90'2014}A. Puglisi, S. Plumari, and V. Greco, \href{https://journals.aps.org/prd/abstract/10.1103/PhysRevD.90.114009}{Phys. Rev. D \textbf{90} (2014) 114009}.

\bibitem{Greif:PRD90'2014}M. Greif, I. Bouras, C. Greiner, and Z. Xu, \href{https://journals.aps.org/prd/abstract/10.1103/PhysRevD.90.094014}{Phys. Rev. D \textbf{90} (2014) no. 9, 094014}. 

\bibitem{Hattori:PRD96'2017}K. Hattori, X. G. Huang, D. H. Rischke, and D. Satow, \href{https://doi.org/10.1103/PhysRevD.96.094009}{Phys. Rev. D \textbf{96}, 094009 (2017).}

\bibitem{Kharzeev:Nucl.Phys.A803'2008}D. E. Kharzeev, L. D. McLerran, and H. J. Warringa, \href{https://doi.org/10.1016/j.nuclphysa.2008.02.298}{Nucl.Phys. A \textbf{803}, 227 (2008).}

\bibitem{Fukushima:PRD78'2008}K. Fukushima, D. E. Kharzeev, and H. J. Warringa, \href{https://journals.aps.org/prd/abstract/10.1103/PhysRevD.78.074033}{Phys. Rev. D \textbf{78}, 074033 (2008).}

\bibitem{Kharzeev:Ann.Phys.325'2010}D. E. Kharzeev, \href{https://www.sciencedirect.com/science/article/pii/S0003491609002206?via%3Dihub}{Ann. Phys. (Amsterdam) \textbf{325}, 205 (2010).}
	
\bibitem{Shovkovy:LecNotesPhys871'2013}I. A. Shovkovy, \href{https://link.springer.com/book/10.1007%2F978-3-642-37305-3}{Lect. Notes Phys. \textbf{871}, 13 (2013)}.		
		
\bibitem{Kharzeev:PRD83'2011}D. E. Kharzeev and H. U. Yee, \href{https://journals.aps.org/prd/abstract/10.1103/PhysRevD.83.085007}{Phys. Rev. D \textbf{83}, 085007 (2011).}
		
\bibitem{Braguta:PRD89'2014}V. Braguta, M. N. Chernodub, V. A. Goy, K. Landsteiner,
A. V. Molochkov, and M. I. Polikarpov, \href{https://journals.aps.org/prd/abstract/10.1103/PhysRevD.89.074510}{Phys. Rev. D \textbf{89}, 074510 (2014).}
		
\bibitem{Chernodub:PRB89'2014} M. N. Chernodub, A. Cortijo, A. G. Grushin, K.
Landsteiner, and M. A. H. Vozmediano, \href{https://journals.aps.org/prb/abstract/10.1103/PhysRevB.89.081407}{Phys. Rev. B \textbf{89}, 081407 (2014).}

\bibitem{Das:PRD97'2018} A. Das, A. Bandyopadhyay, P. K. Roy, and M. G. Mustafa, \href{https://doi.org/10.1103/PhysRevD.97.034024}{Phys. Rev. D \textbf{97}, 034024 (2018).}


\bibitem{Panday:PRD105'2022}P. Panday and B. K. Patra, \href{https://journals.aps.org/prd/abstract/10.1103/PhysRevD.105.116009}{Phys. Rev. D \textbf{105}, 116009}
		
\bibitem{Das:PRD101'2020}A. Das, H. Mishra, and R. K. Mohapatra, \href{https://journals.aps.org/prd/abstract/10.1103/PhysRevD.101.034027}{Phys. Rev. D \textbf{101},
034027 (2020).}
		
\bibitem{Thakur:PRD100'2019} L. Thakur and P. K. Srivastava, \href{https://journals.aps.org/prd/abstract/10.1103/PhysRevD.100.076016}{Phys. Rev. D \textbf{100}, 076016 (2019).}
		
\bibitem{Feng:PRD96'2017}B. Feng, \href{https://journals.aps.org/prd/abstract/10.1103/PhysRevD.96.036009}{Phys. Rev. D \textbf{96}, 036009 (2017).}
		
\bibitem{Rath:EPJC82'2022}S. Rath and S. Dash, \href{https://link.springer.com/article/10.1140/epjc/s10052-022-10757-4}{Eur. Phys. J. C \textbf{82}, 797 (2022).}
		
\bibitem{Shaikh}A. Shaikh, S. Rath, S. Dash and B. Panda, \href{https://arxiv.org/abs/2210.15388}{arXiv:2210.15388}
		
\bibitem{Schenke:PRL108'2012}B. Schenke, P. Tribedy, and R. Venugopalan,
\href{https://journals.aps.org/prl/abstract/10.1103/PhysRevLett.108.252301}{Phys. Rev. Lett. \textbf{108}, 252301 (2012).}

\bibitem{Bhatt:PRD99'2018}J. R. Bhatt, A. Das, and H. Mishra \href{https://journals.aps.org/prd/abstract/10.1103/PhysRevD.99.014015}{Phys. Rev. D \textbf{99}, 014015 (2018)}.

\bibitem{Das:PRD102'2020}A. Das, H. Mishra, and R. K. Mohapatra, \href{https://journals.aps.org/prd/abstract/10.1103/PhysRevD.102.014030}{Phys. Rev. D \textbf{102}, 014030 (2020)}.

\bibitem{Dey:PRD102'2020}D. Dey and B. K. Patra, \href{https://journals.aps.org/prd/abstract/10.1103/PhysRevD.102.096011}{Phys. Rev. D \textbf{102}, 096011 (2020).}

\bibitem{Dey:PRD104'2021}D. Dey and B. K. Patra, \href{https://journals.aps.org/prd/abstract/10.1103/PhysRevD.104.076021}{Phys. Rev. D \textbf{104}, 076021 (2021)}

\bibitem{Kurian:PRD103'2021}M. Kurian, \href{https://journals.aps.org/prd/abstract/10.1103/PhysRevD.103.054024}{Phys. Rev. D \textbf{103}, 054024 (2021).}

\bibitem{Zhang:EPJC81'2021} H. Zhang, J. Kang, and B. Zhang, \href{https://link.springer.com/article/10.1140/epjc/s10052-021-09409-w}{Eur. Phys. J. C \textbf{81}, 623 (2021).}

\bibitem{Khan:arxiv}S. A. Khan and B. K. Patra, \href{https://arxiv.org/abs/2211.10779}{arXiv:2211.10779 }

\bibitem{Fukushima:PLB591'2004} K. Fukushima, \href{https://doi.org/10.1016/j.physletb.2004.04.027}{Phys. Lett. B \textbf{591}, 277 (2004).}

\bibitem{Ghosh:PRD73'2006} S. K. Ghosh, T. K. Mukherjee, M. G. Mustafa, and R. Ray,
\href{https://doi.org/10.1103/PhysRevD.73.114007}{Phys. Rev. D \textbf{73}, 114007 (2006).}

\bibitem{Abuki:PLB676'2009}H. Abuki and K. Fukushima, \href{https://doi.org/10.1016/j.physletb.2009.04.078}{Phys. Lett. B \textbf{676}, 57 (2009).}

\bibitem{Tsai:JPG36'2009}
H. M. Tsai, and B. Muller, \href{https://doi.org/10.1140/epja/i2016-16222-y}{J. Phys. G \textbf{36}, 075101 (2009).}

\bibitem{Chandra:PRC76'2007} V. Chandra, R. Kumar  and V Ravishankar, \href{https://doi.org/10.1103/PhysRevC.76.054909}{Phys. Rev. C, \textbf{76}, 054909 (2007).} 

\bibitem{Su:PRL114'2015} N. Su and K. Tywoniuk, \href{https://doi.org/10.1016/j.nuclphysbps.2016.05.063}{Phys. Rev. Lett. \textbf{114}, 161601 (2015).}

\bibitem{Florkowski:PRC94'2016} W. Florkowski, R. Ryblewski, N. Su, and K. Tywoniuk, \href{https://doi.org/10.1103/PhysRevC.97.024915}{Phys. Rev. C \textbf{94}, 044904 (2016).}

\bibitem{Jaiswal:PLB11'2020} A. Jaiswal and N. Haque, \href{https://doi.org/10.1016/j.physletb.2020.135936}{Phys. Lett. B \textbf{811}, 135936 (2020).}

\bibitem{Bannur:JHEP09'2007} V. M. Bannur, \href{https://dx.doi.org/10.1088/1126-6708/2007/09/046}{J. High Energy Phys. \textbf{09} (2007) 046.}

\bibitem{Braaten:PRD45'1992} E. Braaten and R. D. Pisarski, \href{https://doi.org/10.1103/PhysRevD.45.R1827}{Phys. Rev. D \textbf{45}, R1827 (1992).}

\bibitem{Peshier:PRD66'2002}A. Peshier, B. Kämpfer, and G. Soff, \href{https://doi.org/10.1103/PhysRevD.66.094003}{Phys. Rev. D \textbf{66}, 094003 (2002).}
 
 \bibitem{Ayala:PRD98'2018} A. Ayala, C. A. Dominguez, S. Hernandez-Ortiz, L. A. Hernandez, M. Loewe, D. Manreza Paret, and R. Zamora,  \href{https://doi.org/10.1103/PhysRevD.98.031501}{Phys. Rev. D \textbf{98}, 031501(R) (2018).}
 
 
\bibitem{Callen1960}H. B. Callen, \textit{Thermodynamics} (Wiley, New York, 1960).

\bibitem{Scheidemantel:PRB68'2003}T. J. Scheidemantel, C. Ambrosch-Draxi, T. Thonhauser, J. V. Badding, and J. O. Sofo, \href{https://journals.aps.org/prb/abstract/10.1103/PhysRevB.68.125210}{Phys. Rev. B \textbf{68}, 125210 (2003)}.

\bibitem{Bazow:PRC90'2014}D. Bazow and U. Heinz, \href{http://dx.doi.org/10.1103/PhysRevC.90.054910}{Phys. Rev. C \textbf{90}, 054910 (2014).}

\bibitem{Hosoya:NPB250'1985} A. Hosoya and K. Kajantie, \href{https://doi.org/10.1016/0550-3213(85)90499-7}{Nucl. Phys. B\textbf{250}, 666 (1985).}

\end{thebibliography}
\end{document}